# Faulting of rocks at three-dimensional stress field by micro-anticracks

H.O. GHAFFARI[1] , M.H.B.NASSERI [1] & R.PAUL YOUNG [1]

(a)E-mail: h.o.ghaffari@gmail.com

[1] Department of Civil Engineering and Lassonde Institute, University of Toronto, Toronto, 170 college street ,M5S3E3,ON, Canada

**Nucleation of a shear fault is the result of interaction and coalescence of cracks. A three-dimensional (3D) stress state where $\sigma_1 > \sigma_2 > \sigma_3$ generally leads to growth of polymodal faults in brittle rocks with interactions of cracks preferentially aligned parallel to $\sigma_1 - \sigma_2$ plane. Here we report an experimental investigation of the multi-stationary acoustic waveforms from the orthorhombic faulting patterns in sandstone under 3D-polyaxial stress fields. We show that emitted waveforms from micro-cracks in true three-dimensional stress state carry shorter rapid slip phase, which is associated with unconventional micro-cracking developed in maximum and minimum ($\sigma_1 - \sigma_3$) stress plane. Our results implicate that three-dimensional (3D) stress state can quicken dynamic weakening about two times than regular events due to inducing unconventional rupture propagation. These results suggest a new nucleation mechanism of 3D-faulting in upper crust rocks which has important implications for earthquakes instabilities, as well as studies of avalanches associated with dislocations.**

The main mechanisms of faulting support two main macro-scale theoretical models, namely andersonian [1] and slip models [2] which are compatible with field and experimental observations [3-5]. While the classic andersonian model (based on the Coulomb-Mohr criterion) predicts bimodal of faulting, the slip model explains the polymodel patterns [2,6-7], obtained under a true triaxial stress state ($\sigma_1 > \sigma_2 > \sigma_3$). From a micro-mechanical point of view, based on acoustic emission patterns in conventional triaxial testes (i.e., $\sigma_1 > \sigma_2 = \sigma_3$) , the emergence of composite shear fracture leading to faulting is related to the interaction and coalescence of brittle cracks [8-10]. The prevailing perspective regarding the orientation of regular micro cracks is that they preferentially align parallel to $\sigma_1 - \sigma_2$ plane [6,8 ,10]. However, for deep-earthquakes under high pressure-high temperature, the orientation of micro-cracks forming final fault plane possibly is perpendicular to maximum stress state (i.e., anticracks) [11-12]. The main feature of deep-earthquakes with possible anti-crack's sources is that many deep events (deep-focus earthquakes) start up significantly faster than do intermediate or shallow events, where the entire rupture time of deep-earthquakes is about half of shallower events [13-14]. Here, we show that this is not the only case in deep-focus earthquakes and can be observed in shallow earthquakes under true triaxial tests. We report the real-time evolution of 3D faulting patterns leading to Polymodal complex faulting systems. While we visualize the dynamic evolution of ruptures-forming final fault system, under true triaxial test (TTT), we speculate that the nature of earthquakes under TTT can be different from conventional triaxial tests (CTTs). We show that unconventional fault-nucleation in TTTs is associated with the observation of shorter generic phases of emitted waveforms which imprints unique signatures in appropriate phase spaces, different from conventional rupture fronts. Our experiments further implicate a new weakening mechanism so that micro faults reach about two times faster to steady state of their resistance against motion.

Our experimental system is schematically shown in Fig.1a. A cubic sample of Fontainebleau sandstone ($80 \times 80 \times 80mm$) –supported with 3 acoustic emission transducers per each face- is loaded where $\sigma_2 = 35MPa$ and $\sigma_3 = 5MPa$ are fixed during the experiment time (section1 and 4 of supplementary materials for other tests). The superimposed best-located events on the reconstructed micro X-ray patterns showed that the evolution of faulting follows "diffusive" fracture patterns , forming the final shape of fault's patches (Figure1c,d-Fig.S.1). This is a typical example of composite self-organized cracks, observed frequently in conventional cylindrical tests [16, 17]. Despite this similarity, we find out the details of the evolution of acoustic signals corresponding to rupture fronts- are different in TTTs. To test this hypothesis, we use a proposed functional network theory on the recorded multi-stationary acoustic signals (supplementary materials, [18]). Using this theory on multi-stationary acoustic signals ,we found three generic time scales, corresponding to the following phases in Q-profiles or modularity of corresponding functional networks (see Methods part for definition) [18] : (1) nucleation and main deformation phase (2) fast-slip and (3) slow slip or attenuation stage (Fig.S2b). To get use of this recent finding and connect to our tests, first we show that the duration of fast-slip phase (the second evolutionary phase) in CTTs is not sensitive to the confinement pressure nor is it sensitive to the geometry of the sample. In supplementary materials, we have shown that under different loading conditions including different strain rates, acoustic signal feed-back control test, smooth and rough fault frictional interface with 150 and 200 MPa confinement pressures over Westerly granite samples and double-direct shear test with quartz-gouge materials, does not effectively change the duration of the fast-slipping phase of precursor events. This evidence strongly shows that the fast-slip phase is a generic phase, which is not sensitive to any of the variations of loading conditions in CTT or conventional friction tests. We note that the nearly invariant time scale of the second phase has been reported recently [19] on engineering polymeric materials (and interfaces) and has been shown to be insensitive to the dimension of loaded samples. We propose this generic phase is directly linked to fracture surface energy and then any change in energy dissipation sources [20] will alter the duration of the second phase (Fig.S.3).

Interestingly, examining the same interval for true triaxial test shows a shorter time interval (~10μs) for the same sandstone (Fig.2a (and inset), b). We have observed the mentioned phenomenon from different events, occurring in different times and positions. This observation implies a different pulse shape for acoustic-crackling noises under 3D boundary conditions while a universal left-asymmetric shape of the pulse is maintained for both CTT and TTT (section3 of supplementary materials). To further analysis of the shorter fast-slip phase, we use a simple effective temperature model ([19, 21] -section2 of supplementary materials) . The model estimates that the cracks under the triaxial stress field carry a smaller amount of energy (at least 40% of CTT), indicating different nature of cracking. We support the idea of unconventional micro-cracks with observation of micro-cracks activity mainly in maximum and minimum stress-plane, forming possible quasi-anitcracks (section4 of supplementary materials - Fig.S.16). For regular micro-cracks which generally propagate in maximum and intermediate stress-planes, the duration of the fast slip phase approaches to CTTs or non-true trixial results (such as our FTB3- section 4 and 5 of supplementary materials). Then, we speculate that unconventional directionality of rupture fronts –induced by far field stress state –results faster slipping period. To see how unconventional cracking change the shape of waveforms ,we examined typical waveforms during the formation of anti-cracks in a multi-anvil test (High pressure –High temperature: HP-HT) on the Olivine samples (section 5 of supplementary materials - Fig.S.16), which revealed ~8μs of the fast-slip phase. This observation on events from HP-HT experiment not only confirm fast rising time of natural deep-focus earthquakes [13-14], also indicates that the origin of the shorter generic phases and "fast-death" of rupture fronts in TTTs is due to irregular cracking. Interactions of such quasi-anticracks eventually form the final fault surface which can be explained with microcrack interaction model such as the three-dimensional solution of Eshelby [6]. The concept of anti-cracks previously has been also used to analyze the failure process in high porosity sandstone as a theory of compaction bands [22], to characterize the triggering mechanism of snow avalanches [23] as well as the localized dissolution of limestone (24). A natural question is whether source of unconventional events are due to nucleation of compaction band in sandstone [22]. To test this approach, we analyzed events from a compaction band test on Diemelstadt Sandstone by a conventional cylindrical geometry set-up (Fig.S.21). Our analysis did not result any short-fast slip phase from the recorded rupture fronts while the patterns of macro- compaction bands through the sample were clear. We conclude that anisotropic propagation of rupture fronts and possible substantially lower fracture surface energy are the main mechanisms of our observations, resulting two times faster weakening rate of micro-cracks. Due to the frequent observation of micro-faults with dominant double couple source mechanism (i.e., shear component),we propose that true 3D stress state can induce rapid weakening on 3D-faults, faster than conventional weakening mechanism(s). This fast-release of energy is associated with fast-acceleration [25] which shortens slip-weakening distance. The nature of this slip-acceleration is different from one demonstrated in [25] which is associated with fault wearing and gouge formation; here we assign the quickened phase to grow of abnormal oblique (micro) cracks.

Furthermore, we show that network-phase spaces of unconventional events, collectively, are different from regular events. Considering mean-temporal of maximum laplacian ($\bar{\lambda}_{\max}$) and spatio-temporal mean of centrality ($\overline{\log < B.C >}$) on networks (see Methods part) , we find that recorded events from TTTs and CTTs collapse in $\bar{\lambda}_{\max} - \overline{\log < B.C >}^{-1}$ plane (Fig.3a), which confirms the recent similar observations from the rock-frictional interfaces (18-Fig.S5,Fig.S.11and Fig.S.14). As we have shown in Fig.3a-inset (and supplementary materials), the conventional tests imprint three main cluster of events in the aforementioned plane, corresponding to energy spectrum of ruptures. Comparing two phase spaces ,we find out that a main difference is the cut-offs from the upper and lower bounds in TTT plane , indicating that probability of finding "abnormal" events in terms of rupture's speed regimes under true triaxial boundary conditions is rare. This indicates that the shapes of wave-forms are also bounded rather than CTT cracking noises (Fig.S.12-13). We conjecture that increasing the slope of the aforementioned phase space reduces "mobility" of the events, inducing weaker events in terms of their energy. To complete our analysis, we speculate that the widely discussed failure criterion of materials under TTTs [2, 26-28] can be inspected in the scale of tiny-amplified events. To envisage such micro-failure criterion, we search for possible relation between independent layers of functional multiplex acoustic networks (Methods part), in which an independent network of sensors is assigned for each main direction(X,Y and Z) . Then, for each occurred rupture front and for each sub-network (layer) , we calculate the reciprocal of the minimum modularity index, namely "maximum resistivity" ( $R_i^I$ indicates the maximum resistance against motion from $i$ th layer or sub-network in the first phase of Q-profiles - Fig.S4 and Fig.S6-S14). Since $R_i^I$ occurs in the main deformation phase of Q-profiles (i.e., first stage), then it is proportional with the maximum strength in that direction. Remarkably, the events forming the Polymodal fault-system collapse in the $J_1$-$J_2$ parameter space in which $J_1 = R_1^I + R_2^I + R_3^I$ and $J_2 = R_1^I R_2^I + R_2^I R_3^I + R_1^I R_3^I$ (Fig.3b), represent an excellent power-law : $J_2 \propto J_1^b, b \approx 2.6$ (also see section3 of the supplementary document for more results). Interestingly, the obtained power-law and similar relations (Fig.3b-inset) on other parameters match with the macro-failure criterion presented by Reches [2], Mogi [26] and Haimson [27], indicating that a similar mechanism governs both precursor and final failures (i.e., significant drop of the driving stress). In other words, we have shown that the long-standing problem of failure-criterion in 3D stress field has its origin in each single recorded precursor event (i.e., local failure). In supplementary materials (section3-Fig.S.10-S.13, and section 5), we discussed the possible connection of local failure metrics and global network metrics.

Our results have significant implications for earthquake seismology, rock-mass stability and fluid migration in fractured rocks. Accommodation of 3D strain due to polymodal fracture patterns change strength and permeability of rock mass. For the first time, we showed oblique faulting can change slip-weakening rate and accelerate the rate of energy release. This implicates a sharper source time function for further modeling of our experiments. Indeed, our results can be assumed as an extension to detachment fronts in micro-faults [19] to a general concept of "quasi-anti rupture fronts". We showed events from deep-focus earthquakes can share some similarity to shallow earthquakes, promoting recent approach on similarity of deep earthquakes to their shallow counterparts [34]. Our results might be applicable for emitted events from dislocations in atomic scales which radiate elastic energy in terms of acoustic emissions [29]. Developing models based on introduced multiplex-functional networks and evaluating other properties of natural earthquakes are our future works.

## Methods

**Summary of Experimental Procedures:** Cubic-samples of the saturated Fontainebleau sandstone are loaded at the University of Toronto, in a home-made true triaxial cell (30). The cell can be used to measure permeability in different directions as well as to measure the heat conductivity of samples. Especial attention was paid to keep the balance of loads in parallel directions to avoid possible torque. The strain rate for all axial directions was $3 \times 10^{-6} s^{-1}$. The waveforms were recorded using 18 piezoceramic transducers at a 10MHz sampling rate while triggered sensors are employed to detect the events a threshold of 60mv and minimum 6. Another system simultaneously and continuously records all noises from all transducers during the experiments. For conventional compressive tests, we used the results of previously reported experiments (31) while the samples are saturated and the events are chosen from precursor rupture fronts. A Fontainebleau sandstone specimen (length = 88 mm, diameter = 40 mm) was deformed inside a triaxial. A network of piezoceramic transducers (PZT) was used in10MHz sampling rate with similar criterion to the Polyaxial test to detect the events. In the supplementary document (section1 and 5) , we have -also-included the summary of other tests (and their multiplex networks attributes) on Westerly Granite samples under different loading conditions and geometry as well as smooth and rough rock-frictional interfaces (17,18). In addition, we analysed multi-stationary waveforms from a Multi-anvil test (High pressure –High temperature tests) on an olivine sample to infer the possible origin of shorter waveforms (section 5-supplemntary material).

**Functional Multiplex Networks on Acoustic Emission Waveforms:** We define a multiplex network on the cubic geometry over acoustic transducers where it includes three main layers (X,Y and Z directions correspond to minimum, intermediate and maximum principle stress as far-field driving stress -field ) . To build networks, we use the algorithm introduced in (18)-also see supplementary material. Two cases were analyzed through this research: global networks or interdependent layers and local-sub networks as independent layers of mathematical graphs. Each node in functional networks was characterized by its degree $k_i$ . For a given network with $N$ nodes, the degree of the node and Laplacian of the connectivity matrix were defined by $k_i = \sum_{j=1}^{N} a_{ij}; L_{ij} = a_{ij} - k_i \delta_{ij}$ where $k_i, a_{ij}, L_{ij}$ are the degree of $i$ th node, elements of a symmetric adjacency matrix, and the network Laplacian matrix, respectively. The eigenvalues $\Lambda_\alpha$ are given by $\sum_{j=1}^{N} L_{ij}\phi_j^{(\alpha)} = \Lambda_\alpha \phi_i^{(\alpha)}$, in which $\phi_i^{(\alpha)}$ is the $i$ th eigenvector of the Laplacian matrix ($\alpha$=1,...,$N$).We define $\lambda_{\max} = (-\Lambda_N)$ as the maximum eigenvalue of the Laplacian of the network. The betweenness centrality (B.C) of a node is defined as (32):

$$B.C_i = \frac{1}{(N-1)(N-2)} \sum_{\substack{h,j \\ h \neq j, h \neq i, j \neq i}}^{N} \frac{\rho_{hj}^{(i)}}{\rho_{hj}} \quad (1)$$

in which $\rho_{hj}$ is the number of the shortest path between $h$ and $j$ ,and $\rho_{hj}^{(i)}$ is the number of the shortest path between $h$ and $j$ that passes $i$. The spatio-temporal average of B.C is indicated by $\overline{\log\langle B.C\rangle}$ where <…> and bar-sign correspond to the spatial (i.e., nodes) and temporal averages, respectively. In (18) , we speculate that events from saw and rough fault cut-experiments flow in $\bar{\lambda}_{\max} - \overline{\log < B.C >}^{-1}$ phase diagram with three disgusted trends, corresponding to rupture regime. Here, an addition to mapping events to mentioned global phase diagram, we map each event into local phase diagrams (i.e., sub-networks' phase spaces): $\bar{\lambda}_{\max}^{x,y,z} - \overline{\log < B.C >_{x,y,z}}^{-1}$ .

The Q-profiles as the networks' modularity characteristic is given by (32-33):

$$Q = \sum_{s=1}^{N_M} [\frac{l_s}{L} - \left(\frac{d_s}{2L}\right)^2], \quad (2)$$

in which $N_M$ is the number of modules (clusters) , $L = \frac{1}{2}\sum_i^N k_i$ , $l_s$ is the number of links in module and $d_s = \sum_i k_i^s$ (the sum of nodes degrees in module s). Using an optimization algorithm (Louvain algorithm-(32)), the cluster with maximum modularity ($Q$) is detected.

## Supplementary Information

Supplementary information accompanies this paper.

**Acknowledgements**

We would like to acknowledge and thank M.G.Tabari , S.D.Goodfellow (University of Toronto) and L. Lombos for technical Laboratory helps and discussion on the results. The main triaxial cell is the hard work of L.Lombos (Ergotech, UK) over years. The first author would like to appreciate from A.Schubnel (Laboratoire de Géologie de l'Ecole normale supérieure,France) , P.Benson (University of Portsmouth) , B.D.Thompson (Mine Design Engineering, Canada), P. Meredith (UCL ,UK) and D.A.Lockner (USGS,US) for supporting the research with providing their data set. Discussion with Z.Reches (University of Oklahoma) was also useful. K.Xia (University of Toronto) kindly encouraged and commented on the results.

**Author Contributions**

All authors contributed to the analysis the results. H.O.G.H. performed the calculations and wrote the manuscript. M.H.B.Nasseri accomplished and designed the tests and R.P.Y. supervised the research, provided the data set and the analysis of the results.

**Competing interests statement**

The authors declare no competing financial interests

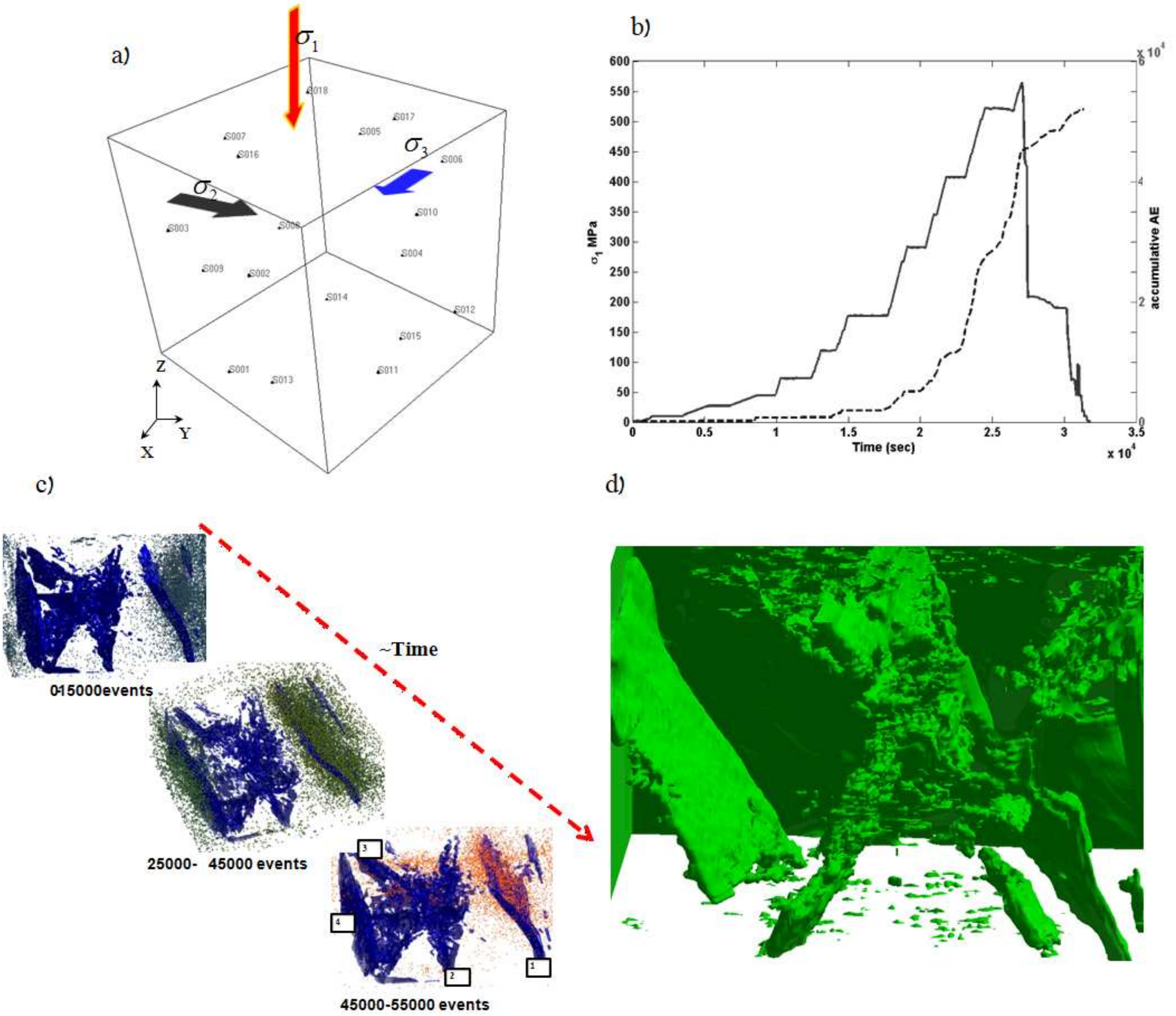

**Fig. 1. A study of Orthorhombic faulting under the true triaxial tests**.(a) a schematic of the experiment procedure and remote stress field stress configuration has been shown where the numbers on each face of the cube shows the piezoelectric transducers.(b) The evolution of the main driving stress field ( $\sigma_1$ ) and accumulated recorded acoustic emissions during a few hours of the experiment . The stairlike trend of the loading curve is due to pausing the test in order to measure the permeability in three directions. (c,d ) Superimposed events on the best-located events on the reconstructed of the X-ray patterns of the fault-system and the temporal/spatial evolution of the events (i.e., event-sets' sequences). The first two event-sets show the slow growth of the external fault-system's patches while the fast-growth of the internal branches of the Orthorhombic faulting system is observed around the main stress drop (Fig.S1). The numbers shows 4 main fault sets. (d) A close view of 3D - fault system .

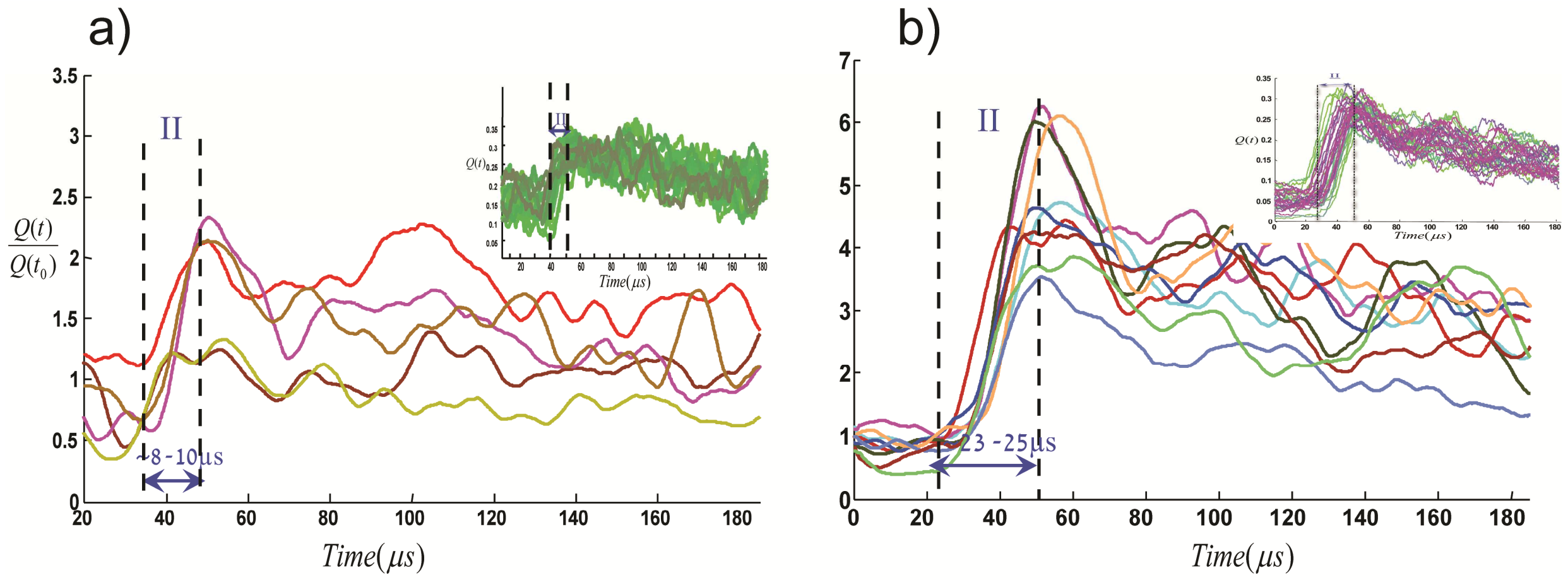

**Fig. 2.Q-profiles and Fast-slip phase in TTT and CTT .** (a) Q-profiles from a TTT test (FTB4) show an ~8-10 μs duration of the fast-slipping phase (inset: 150 events from TTT-FTB2; see section 4 in the supplementary information). (b)The same sandstone under the conventional test shows a longer fast slip phase (~23-25 μs). Inset shows 45 events from the CTT.

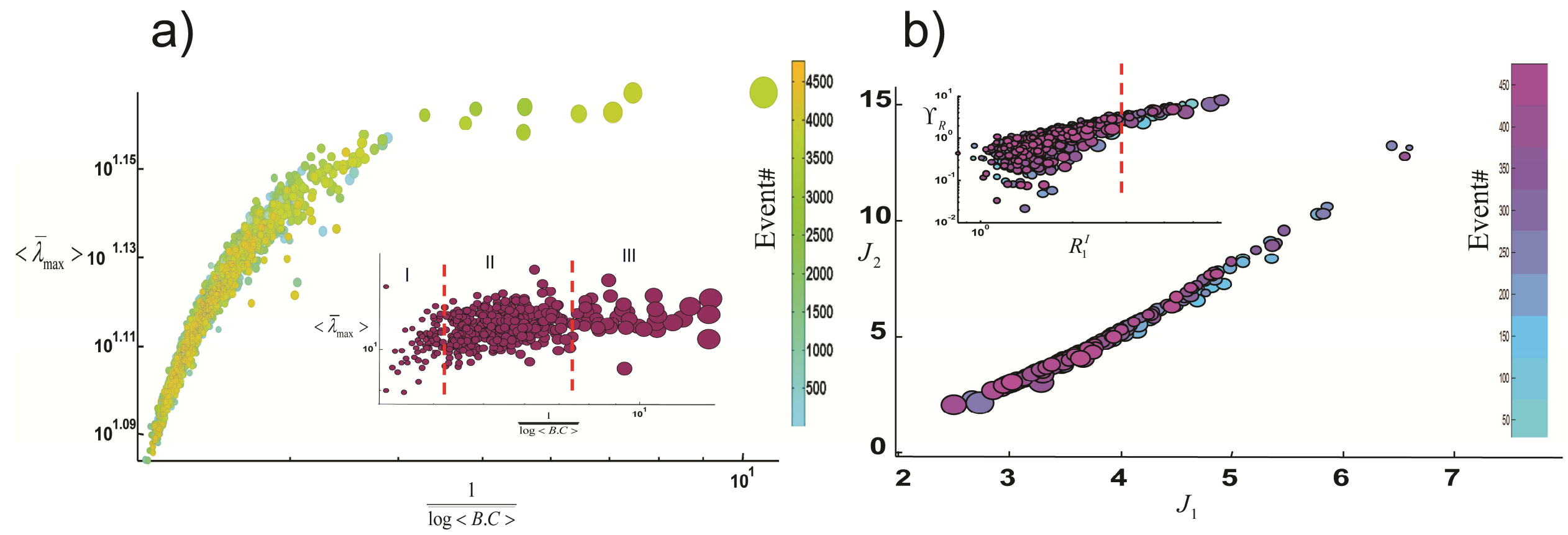

**Fig. 3. Global and local Networks' phase spaces on TTT and CTT** (a) $\bar{\lambda}_{max} - \overline{\log < B.C >}^{-1}$ parameter space on ~4500 events (including weak events) an hour prior to the main-final faulting development in FTB4 (also see section 3 of supplementary materials). Inset shows events from the CTT at the same phase space with three main classified classes corresponding to rupture fronts energies. (b) Sub-network's phase diagrams: Collapsing ~520 best-strong events from the FTB4 test in the $J_1$ - $J_2$ parameter space, where $J_1 = R_1^I + R_2^I + R_3^I$ and $J_2 = R_1^I R_2^I + R_2^I R_3^I + R_1^I R_3^I$. A power law relation satisfies an excellent collapsing of events in $J_1$ - $J_2$ phase diagram as: $J_2 \propto J_1^b, b \approx 2.6$. In the inset, we have shown the $R_1^I - \Upsilon_R$ phase diagram with $\Upsilon_R = \sqrt{(R_1^I - R_2^I)^2 + (R_2^I - R_3^I)^2 + (R_3^I - R_1^I)^2}$ for the same events (Also see Fig.S.5-Fig.S.15). Further analysis of events from rock-frictional interfaces (Fig.S19) indicates that $R_1^I - \Upsilon_R$ phase space covers main features of failures.

# Supplementary Materials

## I) Summary of Experimental Procedures:

Cubic-samples of the saturated Fontainebleau sandstone are loaded at the University of Toronto, in a true triaxial cell [1]. The cell can be used to measure permeability in different directions as well as to measure the heat conductivity of samples. Especial attention was paid to keep the balance of loads in parallel directions to avoid possible torque. The strain rate for all axial directions was $3 \times 10^{-6} s^{-1}$. The waveforms were recorded using 18 piezoceramic transducers at a 10MHz sampling rate while triggered sensors are employed to detect the events a threshold of 60mv and minimum 6. Another system simultaneously and continuously records all noises from all transducers during the experiments. For conventional compressive tests, we used the results of previously reported experiments [2] while the samples are saturated and the events are chosen from precursor rupture fronts. A Fontainebleau sandstone specimen (length = 88 mm, diameter = 40 mm) was deformed inside a triaxial. A network of piezoceramic transducers (PZT) was used in10MHz sampling rate with similar criterion to the Polyaxial test to detect the events. In the supplementary document (section1 and 5) , we have -also- included the summary of other tests (and their multiplex networks attributes) on Westerly Granite samples under different loading conditions and geometry as well as smooth and rough rock-frictional interfaces [7]. In addition, we analysed multi-stationary waveforms from a Multi-anvil test (High pressure –High temperature tests) on an olivine sample to infer the possible origin of shorter waveforms (section 5-supplemntary material).

## II) Functional Multiplex Networks on Acoustic Emission Waveforms

In [7], we used functional networks as a class of recurrence networks to analysis precursor events during the evolution of frictional interfaces. Here, we use a similar algorithm on waveforms from acoustic emissions as our laboratory earthquakes in mesoscales. First we define a global or multiplex network on the cubic geometry where It includes three main layers (X,Y and Z directions correspond to minimum, intermediate and maximum principle stress as far-field driving stress -field ) . The two extreme cases can be inferred [25-26]: (1) independent layers and (2) fully coupled layers. The "local sub-networks" were in the first class and we used them to investigate the attributes of each micro-crack on each layer. To build networks (both global and local) ,we use the following algorithm [7]: **(1)** normalization of waveforms in each station where the maximum amplitude of each recorded waveform –after normalization-is unit. **(2)** Division of each time series to maximum segmentation. Then, we consider each recorded point in each waveform with the length of *T.* The *j*th segment from *i*th time series ($1 \le i \le N$) is denoted by $x^{i,j}(t)$. We put the length of each segment as unit. This essentiality considers the high temporal- resolution of the systems' evolution. **(3)** Adding edges: $x^{i,j}(t)$ is compared with $x^{k,j}(t)$ to create an edge among the aforementioned segments. If $d(x^{i,j}(t),x^{k,j}(t)) = \|x^{i,j}(t) - x^{k,j}(t)\| \le \xi$ , we set $a_{ik}(j) = 1$ otherwise $a_{ik}(j) = 0$ in which $a_{ik}(j)$ is the component of the connectivity matrix. **(4)** Threshold level ($\xi$): To select a threshold level, we use betweenness centrality (B.C) ,which can be assumed as a measure of "load" on each node. The details of the method have been explained in [27-28], and it has been proven that using this method quantitatively is equal to using edge density. **(5)** Increase the resolution of visualization : To decrease the sensitivity of the networks and knowing that recurrence networks–generally- reveal a good performance in a small number of nodes, we increased the size of the adjacency matrix with the simple interpolation of *d* using cubic spline interpolation. Next, to set-up local networks (or sub-networks), for each parallel pair face in a cubic sample, a similar procedure was repeated to form a network. The only change lay in increasing the size of the adjacency matrix with simple interpolation of $d_{sub\text{-}network}$ using cubic spline interpolation which becomes equal to the global network size (i.e., number of nodes). Then for both networks, the numbers of nodes were 18 and we kept the number of nodes as the con-stant value. We also used some main networks' metrics. Each node was characterized by its degree $k_i$ and the clustering coefficient. For a given network with $N$ nodes, the degree of the node and Laplacian of the connectivity matrix were defined by $k_i = \sum_{j=1}^{N} a_{ij}; L_{ij} = a_{ij} - k_i \delta_{ij}$ where $k_i, a_{ij}, L_{ij}$ are the degree of $i$ th node, elements of a symmetric adjacency matrix, and the network Laplacian matrix, respectively. The eigenvalues $\Lambda_\alpha$ are given by $\sum_{j=1}^{N} L_{ij} \phi_j^{(\alpha)} = \Lambda_\alpha \phi_i^{(\alpha)}$, in which $\phi_i^{(\alpha)}$ is the $i$ th eigenvector of the Laplacian matrix ($\alpha = 1,...,N$).. With sorting the indices $\{\alpha\}$ in decreasing order of the eigenvalues, we have: $0 = \Lambda_1 \geq \Lambda_2 \geq ... \geq \Lambda_N$ and we define $\lambda_{\max} = (-\Lambda_N)$ as the maximum eigenvalue of the Laplacian of the network. We also used the betweenness centrality (B.C) of a node as the measure of "load" [29]:

$$B.C_i = \frac{1}{(N-1)(N-2)} \sum_{\substack{h,j \\ h \neq j, h \neq i, j \neq i}}^{N} \frac{\rho_{hj}^{(i)}}{\rho_{hj}} \qquad (1)$$

in which $\rho_{hj}$ is the number of the shortest path between $h$ and $j$ ,and $\rho_{hj}^{(i)}$ is the number of the shortest path between $h$ and $j$ that passes $i$. The spatio-temporal average of B.C is indicated by $\overline{\log\langle B.C\rangle}$ where <...> and bar-sign correspond to the spatial (i.e., nodes) and temporal averages, respectively. In [7], we speculate that events from saw and rough fault cut-experiments flow in $\overline{\lambda}_{\max} - \overline{\log < B.C >}^{-1}$ phase diagram with three disgusted trends, corresponding to rupture regime. Here, an addition to mapping events to mentioned global phase diagram, we map each event into local phase diagrams (i.e., sub-networks' phase spaces): $\overline{\lambda}_{\max}^{x,y,z} - \overline{\log < B.C >_{x,y,z}}^{-1}$ .

In [7], we found out universal pattern of Q-profiles (i.e., dynamic evolutionary in each recorded event) over the acoustic emissions emitted during evolution of frictional interfaces (Fig.S.2b). The universality patterns of the phases in the Q-profiles resembled the slip-profiles obtained in PMMA and other studies [6].We also use the average skewness of Q-profiles as the asymmetry measure of pulse shapes which is quantified by [11]:

$$\Sigma = \frac{\frac{1}{T}\int_0^T Q(t)(t-\overline{t})^3 dt}{[\frac{1}{T}\int_0^T Q(t)(t-\overline{t})^2 dt]^{3/2}} \qquad (2)$$

in which $\overline{t} = \frac{1}{T}\int_0^T tQ(t)dt$ and $T = 408\mu s$ (i.e., duration of waveforms). Following [6,7], the first stage in Q-profiles of the functional acoustic-damage networks is the nucleation and the main deformation phase where shows the nature of rupture and energy release in the crack tip. To amplify this regime, we use the reciprocal of normalized (to the rest value) Q-profiles: R-profiles which show the resistance and strength of particles against deformation (Fig.S.3-S.4). For each sub-networks, the maximum value of R-profiles –occurred in the first phase of Q-profiles- is indicated by $R_i^I$ which can be assumed as an indicator of the maximum strength in that direction .

## 1. More features from the experiments

In this supplementary document, we demonstrate more evidence of our claims in the main text. The following is **the list of the experiments that we have used to support our claims**:

a. True triaxial tests (TTT) : Three Fontainebleau rock samples in cubic shapes (FTB2,FTB3,FTB4) were loaded at the home-made cell. In one of the tests ,we did not make a complete balance of loading in X-direction (FTB3- non true triaxial test) .We will show this effect in analysis of the monitored events.
b. Conventional cylindrical geometry of Westerly Granite with different strain rates and geometries including fast-strain rate , AE-feedback control test ,slow strain rate (4).
c. Saw-cut and natural rough Granite-Granite frictional interfaces (7)
d. Double shear experiments with different gouge materials (glass and quartz-sands)(5)
e. Conventional cylindrical geometry of Fontainebleau sandstone rock samples (2)
f. Compaction band test : with Conventional cylindrical geometry of Diemelstadt sandstone rock samples (10)
g. Conventional cylindrical geometry of Basalt (9)
h. Conventional cylindrical geometry of Concrete (Portland cement)(8)
i. High pressure-High Temperature (Multianvil test) on Ge-Olivine

The main focus of this supplementary is on TTTs and in particular FTB4. First, we examine duration of the second phase in Q-profiles in different materials and under different conventional loading conditions. The materials in this section is used to show that unusual rupture fronts in micron and sub-micron scales are not due to geometry of the samples or loading rate or other details of conventional loading conditions. Next, we develop a framework based on R-profiles to extract failure phase diagrams per each recorded events using sub-networks on FTB4 and other CTTs. Furthermore, using global (multiplex network) networks' phase planes and Q-profiles, we illustrate the unconventional nature of cracking in true triaxial tests . At the end, the waveforms from multi-anvil test is analyzed to validate our assumption on unusual characteristics of TTT's rupture fronts.

In Fig. S.1, we have shown more features from FTB4 . We also depicted the sequence of formation of faults' patches in association with stress-time and the second temporal derivation of strains in X,Y and Z directions. For a visualization of the fault-system, the reconstructed images of the stacked digital images (Micro-CT images), were filtered by a Hough transformation to detect features of the images (Fig.S.1.b, d).

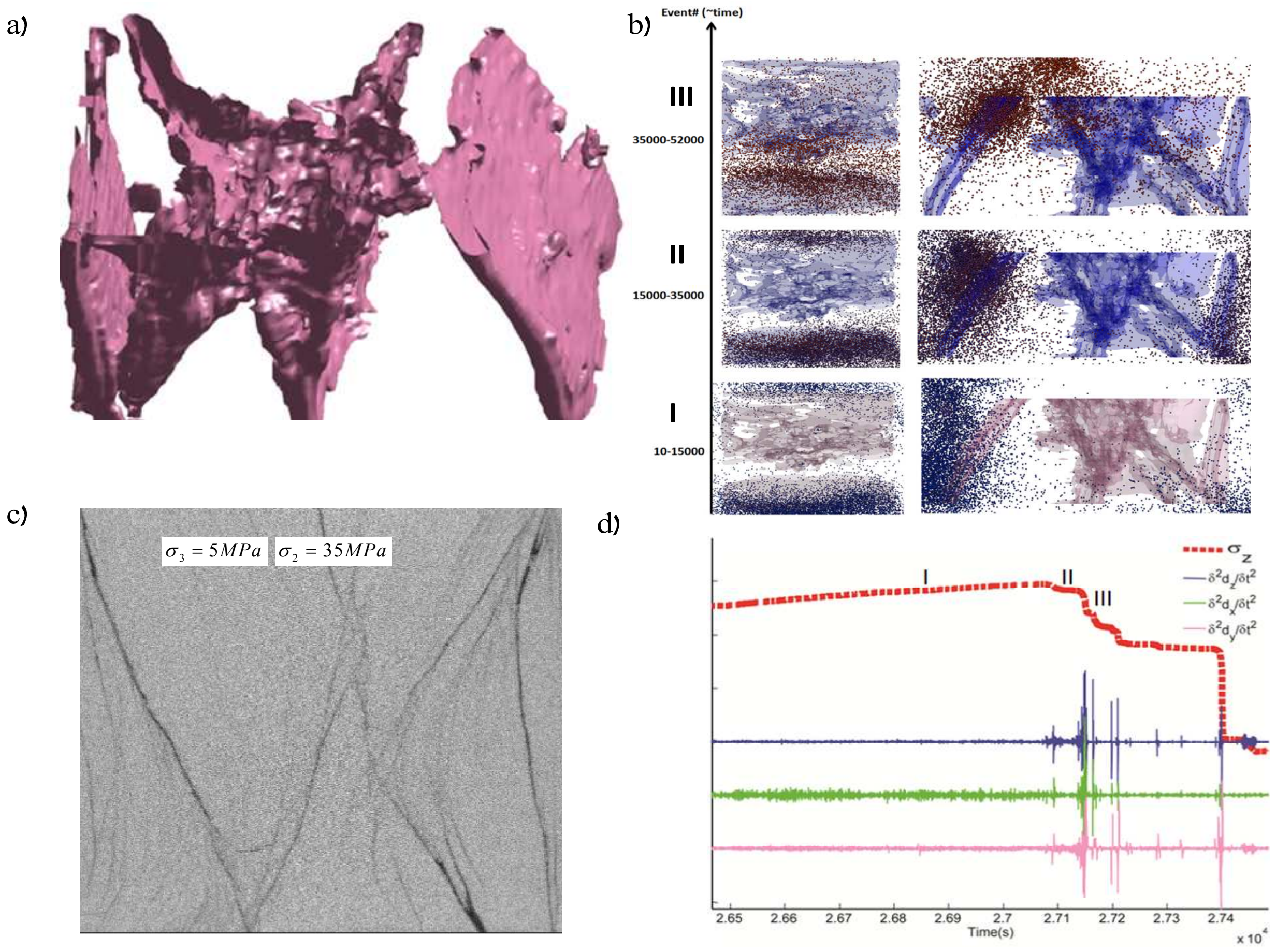

**Figure S.3**. a) Visualization of fault-system in the Polyaxial test-FTB4 shows Orthorhombic faulting system. b) Superimposed events on the best-located events on the reconstructed of the X-ray patterns of the fault-system and the temporal/spatial evolution of the events (i.e., event-sets' sequences). The sequences' events for top (left) and front (right) views have been visualized. c) a cross-section of FTB4 (micro-CT image). (d) Scaled –evolution of the driving stress field ($\sigma_z$) and the second derivation of strain gauges per each direction for FTB4 test. Also, see Fig S.1.e.

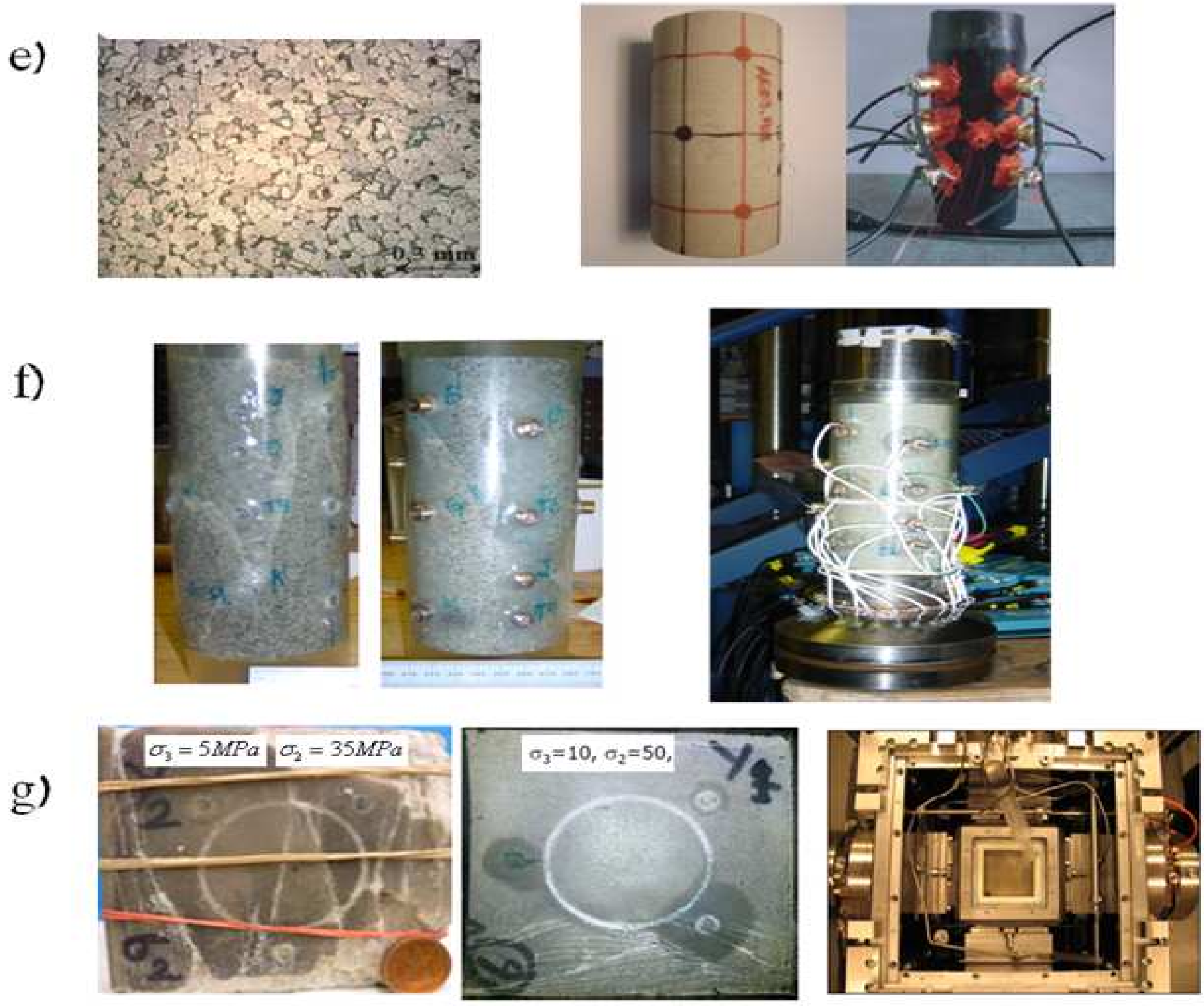

**Figure.S1. (continue)** –e) fine grain Fouintblue sandstone with >90% quartz and the cylindrical sample with mounted piezoelectric sensors (data set and pictures courtesy of Schubnel (2)). f) Westerly Granite samples (see Fig.S.2a-d)-data set Courtesy of B.D.Thompson. g) The final configuration of FTB4 and FTB3 using the home made cell to load and measure permeability (also see (1) for more information on design of the cell).

In FigS.1e-g, we have shown three rock samples (cylindrical Sandstone sample, Westerly Granite ,and Cubic-TTT sample)with acoustic-piezoelectric sensors set-up on them. Next, we prove the following lemma: *Under different loading condition and sample sizes, the fast-slip phase of Q-profiles do not effectively change in conventional compressive tests* (CTT-Fig1S.e) . To test this proposal, we use the results of different loading condition tests on Westerly Granite, reported previously in (4). The sample size for these sets of events was: 190 mm (height) and 76mm in diameter of the cylindrical rock samples. In Fig.S.2, events from different stages of the fast-loading test (with strain rate $10^{-5}/s$ -Fig.S.2 a and c) are compared with events from saw-cut events (Fig.S.2b) and AE-feedback control tests (FigS.2d-see (4) for the details of the test and Fig.S1.f for the final cracks con-figurations and AEs- sensors set-up)). In Fig..S.2.e, the Q-profiles has been shown from the events of the double shear experiment on gouge materials with quartz-sandstone (see the details of the experiment in Ref.5). As we have shown, the second phase shows nearly a constant duration of ~20-25μs.

Then, *the shorter fast slip phase in TTT experiment* *is not* *due to the geometry, loading rate, or other loading conditions of samples*. This generic phase (and other phases extensively discussed in (6-7) ) is due to the released energy from the failure or fast-deformation of an asperity or heterogeneity and is directly related to the physics of rupture in mesoscale deformation. In section 5 and discussion section of this document, we will show that unconventional micro crack nucleation shares an important common feature with anti-cracks (as has been suggested for High-pressure/Temperature tests): *fast-death* of waveforms. In the next section, we will use similar approaches to estimate the amount of energy released in the second phase. In addition to the aforementioned experiments, we compare the Q-profiles from saturated and dry rock samples (Basalt-see the details of the experiments in (9)). In Fig.S.3, we showed the effect of "wet" conditions on the evolution of Q-profiles, indicating that the weak asymmetric Q-profile with fluctuations and the cascade-like nucleation/deformation phase of Q-profiles is the signature of lubrications in meso-scales. It is worth mentioning that the duration of the fast slip stage in the saturated case is longer about ~2-4μs and that results in a slower fast-slip stage and then "smoother" profiles.

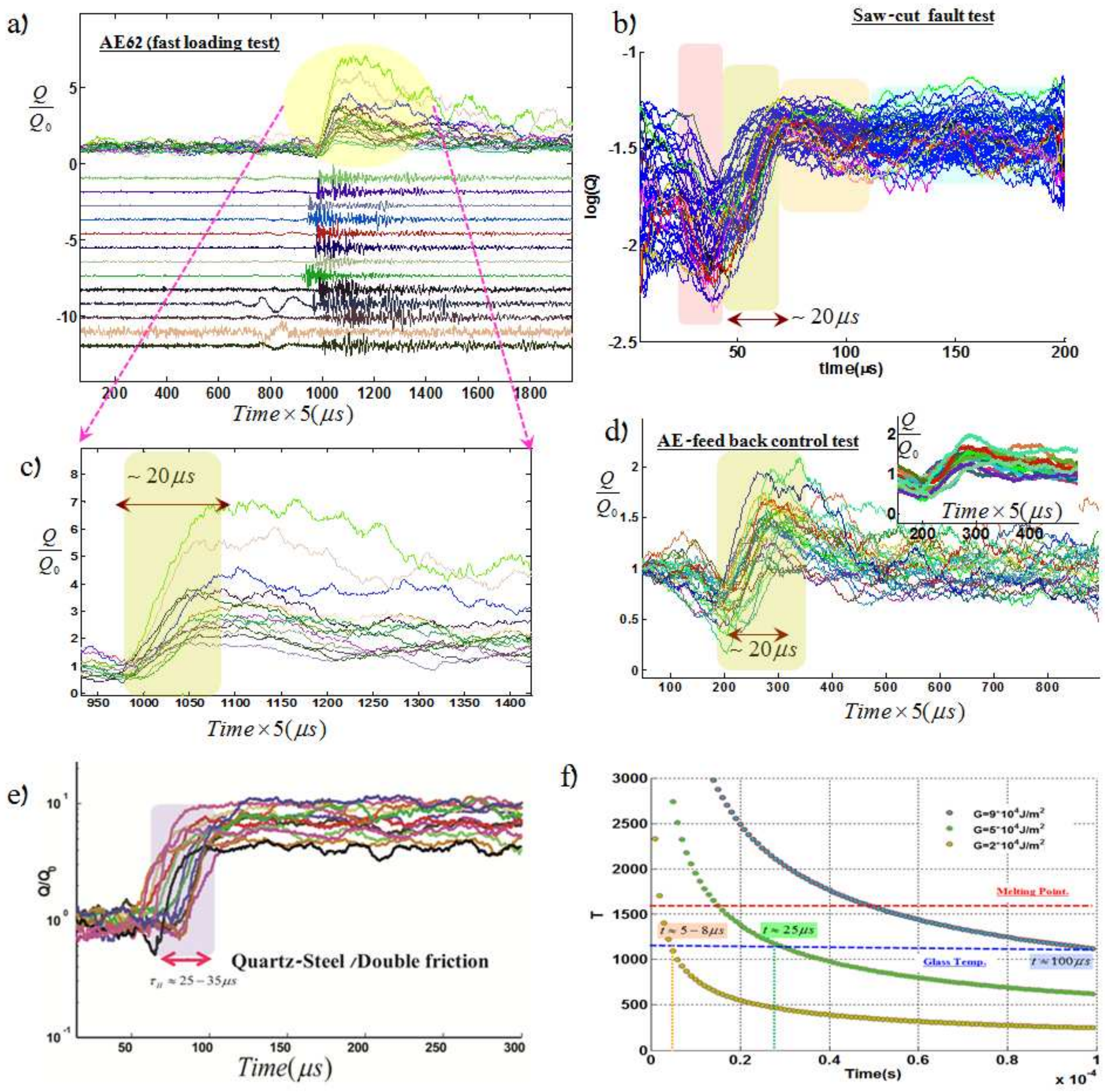

**Figure S.4**. (a-d) ***Loading condition and the geometry of the rock samples do not change the duration of the fast-slip phase in compressive triaxial tests***. We have shown the normalized Q-profiles from different tests on Westerly Granite. (e) Q—profiles from a double-shear test with Quartz-sands as the gouge materials . **Also see Fig.s.21 as the summery of the studied waveforms from different experiments**. (f) an effective temperature model on silicate($SiO_2$) for three different fracture energies .

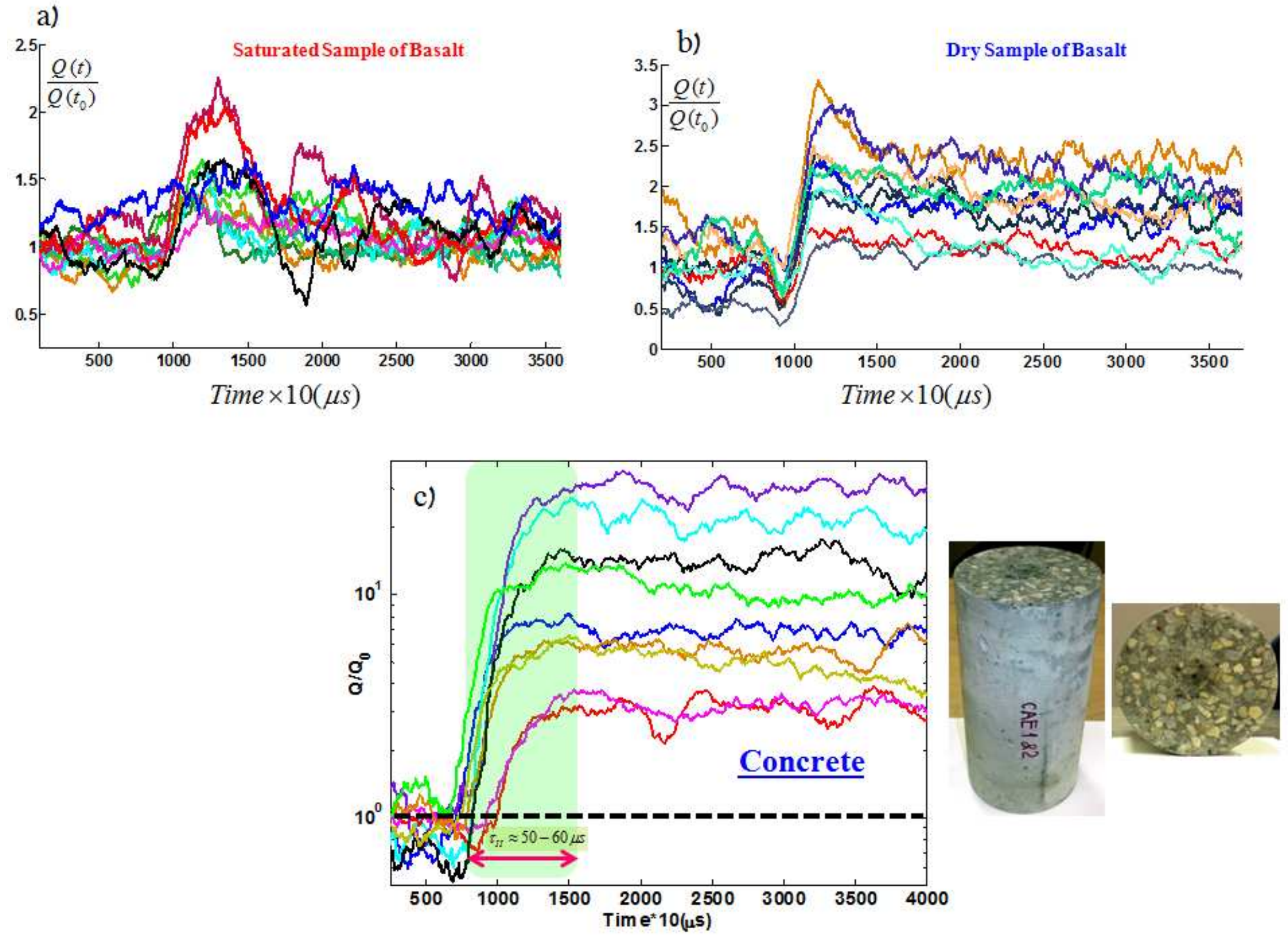

**Figure S.5**. Q-profiles from the Saturated (a) and Dry(b) Basalt samples (9). A soft transition to a fast-slip regime and the less asymmetric shape of the Q-pulses are the most distinguished features of the saturated-events. C) Q-profiles from concrete samples (cemented based materials- Courtesy of original waveforms : Dr. Tatyana Katsaga ; International Journal of Fracture 148.1 (2007): 29-45. ) show a longer period of the second phase indicating a higher toughness and additional energy dissipation mechanisms.

In further analysis on Q-profiles and R-profiles (the methods part in the main text) , in Fig.S.4 , we have compared the R-profiles from TTT and CTT while the details of the evolution of a typical R-profile are shown in Fig.S.4c. The onset of the deformation proceeds with the fast rise in a very short time (~2-4μs) to reach maximum R. We choose this maximum R-value from three sub-networks (corresponding to three main Cartesian directions in cubic-samples) when we are analyzing the sub-networks' metrics. The fast decrease (drop) of R profiles corresponds with the fast-slip phase of Q-profiles; then is longer in CTT 's events.

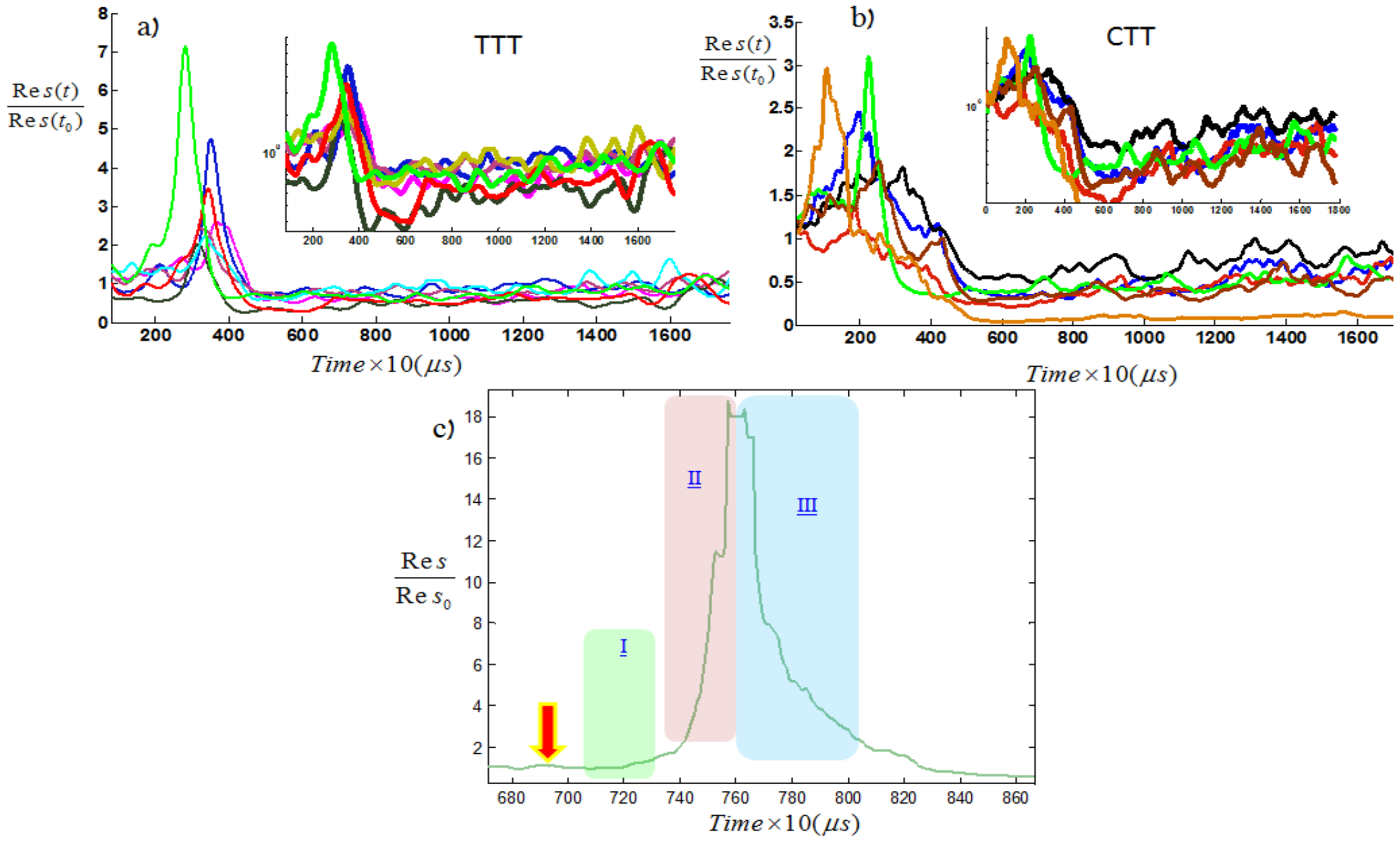

**Figure S.6. (a,b)** Examples of R-profiles from the TTT (FTB4) and the CTT. Inset: semi-logarithmic R-profiles. **(c)** a typical R-profile includes the three main sub-phases: (I) onset of nucleation shown by a red arrow ;(II) fast-rise (~4-8μs) leading to the maximum R and (III) fast-slip part is imprinted as the fast-decay section of R-profile. As It can be followed from the comparison of panel(a) and (b) ,the duration of this phase is longer for CTT.

The evolution of R-profiles (or Q-profiles) in terms of approaching the rest state ( in normalized form to 1), is faster in TTT than CTT, indicating that healing mechanism for this kind of stress regimes is probably faster . The reason can be inferred from the possible correlation of the last evolutionary (dynamic) phase of Q-profiles with the growth of contact areas as has been shown in (6). Then, not only is the fast-slip phase shorter, the slow-slip is shorter and the onset of healing occurs earlier than in other, simpler tests. Another implication of these results can be followed in at least two times faster weakening rate in shallow earthquakes, resulting a new dynamic weakening mechanism. This mechanism leads to lower faster the resistance against driving forces (such as shear) if we consider an interface under TTT conditions.

Next, we visualize the spatio-temporal average of the global network's metrics (Fig.S.5c-also see (7)).For each best located event, we calculate $\frac{1}{\log < B.C >}$ and $\overline{Q}$. Considering that the large amount of $\frac{1}{\log < B.C >}$ corresponds with high-energy tiny ruptures, we conclude that the internal patches of the Orthorhombic fault-system release higher energy than the external patches. The results do match the $\overline{\lambda}_{\max}$ - $\overline{Q}$ phase diagram of global networks (Fig.S.5a-b). The global network parameter spaces –as have been mentioned in the main text- show a significant difference of trend in comparison with the CTT (or events from the rough frictional interfaces) .

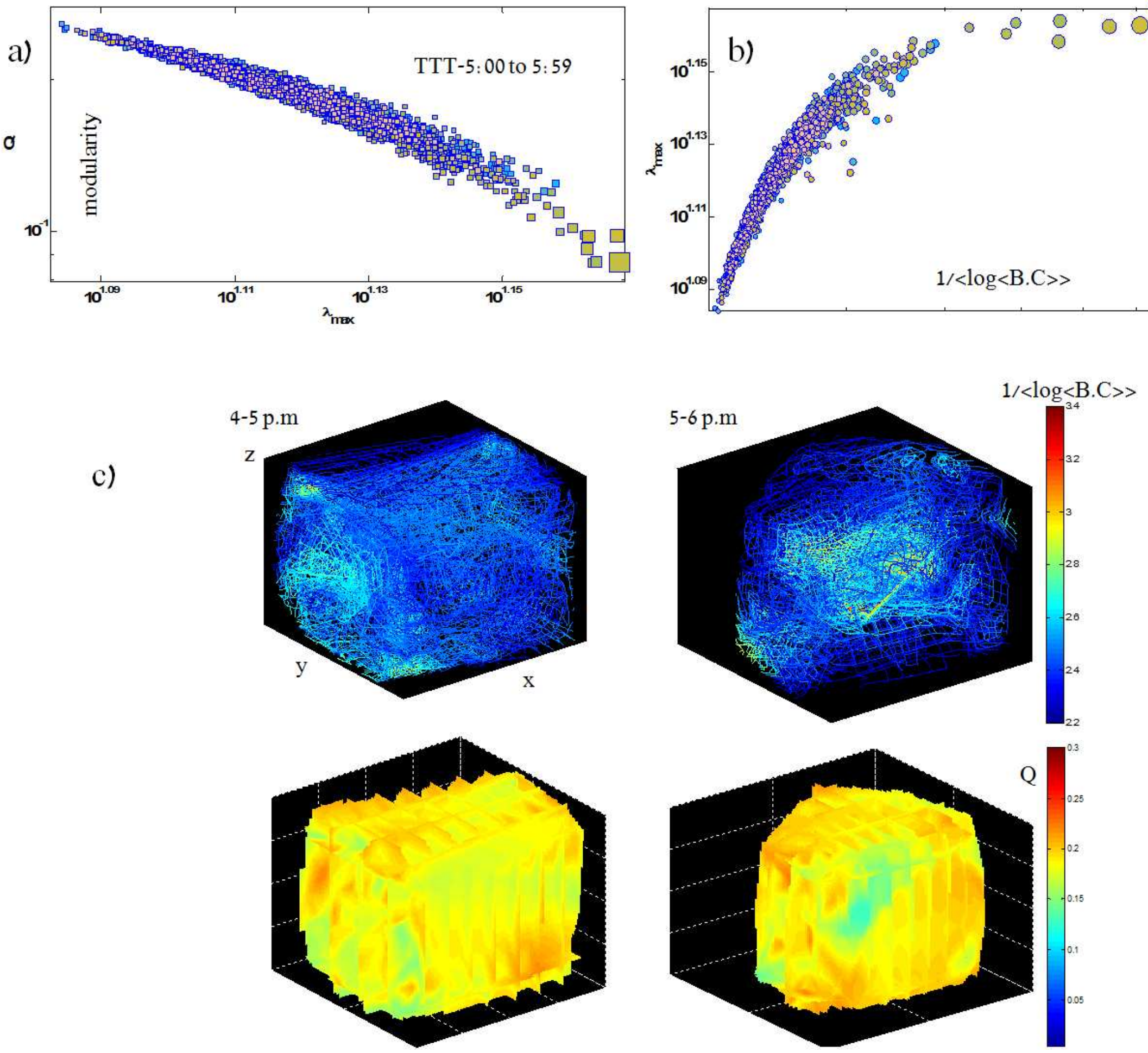

**Figure S.7** Spatial distribution of the temporal average ensemble of network parameters (global network attributes) . (a-b) Global network phase spaces on all detected events (~4500 rupture fronts) in the hour leading to the main stress drop. The size of the circles is proportional with the maximum of $\frac{1}{\log < B.C >}$ . Generally, events with higher energy allocate larger and smaller values of $\frac{1}{\log < B.C >}$ and $\overline{Q}$,respectively. (c) The top-panel shows the $\frac{1}{\log < B.C >}$ during the two hours of the experiment (visualized by 3D contour lines) . Dense activity and high-energy events are formed in the second hour of the experiment while the main failure (and internal patches of M-shape fault system) occurs. Bottom panel: Distribution of $\overline{Q}$ in the second hour of the experiment.

## 2. Calculation on the fast-slip phase of modularity profiles

The model to be used here is a polycrystalline material subject to a simple shear stress. Our key assumption is that generally the motion of a sub-micron crack is governed totally by a thermally activated depinning mechanism. Similar to (6), we propose an effective temperature to explain the observed time scales. As another approximation, because over 90% of Fontainebleau sandstone is quartz minerals, we solve the equation for Silicon dioxide on the scale of sub-microns (or nano scale) while the effect of water is ignored. In other words, we assume that the fast slip regime is approximately the same for both cases(dry and wet). We get used to an approach suggested for the left-hand asymmetric shape of the (average) of avalanches in crackling noise systems (ruptures induced acoustic emissions and Barkhausen noise). Based on this approach (11-12), the asymmetric average shape of the avalanches is due to the role of energy dissipation phenomena (eddy currents and strengthening threshold). Here, we use an equal version for energy dissipation phenomena, originally proposed in (6) and (7) to explain the sudden drop of one of the evolutionary phases (i.e., phase III in Fig.S.2.b). A fast-short time fracturing (phase I) induces a very fast increasing the temperature of a tiny "process zone" which is getting cooled down in a typical time characteristics (i.e., hardening). The main component of the theory is that the fracture energy is encoded in terms of diffusion of heat. The increased temperature with respect to a reference temperature is as follows (6):

$$\Delta T = \frac{-G_t}{4\rho c_p h}[erf(\frac{-h}{\sqrt{4D_T t}}) - erf(\frac{h}{\sqrt{4D_T t}})] \qquad \text{(S1)}$$

in which $erf(x) = \frac{2}{\sqrt{\pi}}\sum_{n=0}^{\infty}\frac{x}{2n+1}\prod_{k=1}^{n}\frac{-x^2}{k} \approx \frac{2}{\sqrt{\pi}}(x - \frac{x^3}{3} + \frac{x^5}{10} - ...)$, $h$ is the thickness of the process zone in which the energy rapidly dissipates, $D_T$ is the thermal diffusivity , $G_t$ is the total energy released in phase I , and $t$ is the cooling time . With first-order approximation, we estimate $t$ as follows:

$$t \approx \frac{G_t^2}{\Theta}, \Theta = 4(\Delta T \rho c_p)^2 \pi D_T \qquad \text{(S2)}$$

in which $\Theta$ is a constant value for a given material. We assume $t \approx t_{II}$.

To proceed furthermore, we plot the variation of effective temperatures for different fracture energies for $SiO_2$ in Fig.S.2f. From Fig.S.2.f we have : $t_{II}^{CTT} \approx 20-25\mu s$ and $t_{II}^{TTT} \approx 8-10\mu s$ (For FTB4- see also section 4) ,which results in $G_{TTT}^{t} / G_{CTT}^{t} \approx .4-.6$. We note that increasing the rate of fast-slip (shorter $t$) phase corresponds with the smaller encoded energy at the second phase. When considering the invariant nature of the second phase under different confinement loadings in CTTs (and other simple tests), we infer that shorter generic stages for the same materials are due to different substantial nature of cracking in meso-scales (and sources of energy dissipation). Since the reported phase – also-is observed in different stages of the test (while source mechanism changes), then we cannot link to the source mechanism (obtained from P-wave arrival). Interestingly, we find out that the sec-ond time interval is longer for concretes –as the cement based materials-due to the cement component and Calcium-Silicate-Hydrate (C-S-H) structures (Fig.S.3c). This is due to additional sources of energy dissipation rather than just creating new surfaces (Buehler&Keten-Rev.Mod.Phys 2010). Also we find out this phase is about ~70µs for TEFLON (as a polymer-amorphous material), close to ~60µs for PMMA (6). Then, tougher materials imprint a longer fast-slip phase which proves the aforementioned approximation. With canceling or suppressing the sources of energy dissipation, we reach the shorter stage. We believe that in our TTTs, this phenomenon does occur, coinciding with the sort of abnormal cracking propagation in mesoscale.
In section 4, we find a link between the micro-cracks' activity in each layer and the duration of the second phase, supporting our approach to the irregularity of caking in some of our TTTs. This violates the conventional perspective on shallow earthquakes which assume earthquakes occur by ordinary fracturing developing in maximum and intermediate principle stress planes. Another implication is that we could find a similar "abnormal" cracking source in other experiments (see the discussion section and results from the multi-anvil test for typical anti-cracking).

## 3. Studies on the functional Sub-Networks (independent layers)

In Fig.S.6, we have shown the procedure to establish sub-networks or layered layers (as a category of Multiplex networks). Each direction (X,Y and Z) is an independent layer with N=6 nodes (extended to 18 nodes –see Methods section). We only consider the edges within the layers and the links between layers are not considered. Fig.S.6c shows an example of the R-profiles from each layer. Further study of each layer's network metrics indicates some interesting features in terms of mechanical interpretation of rupture properties in that layer.

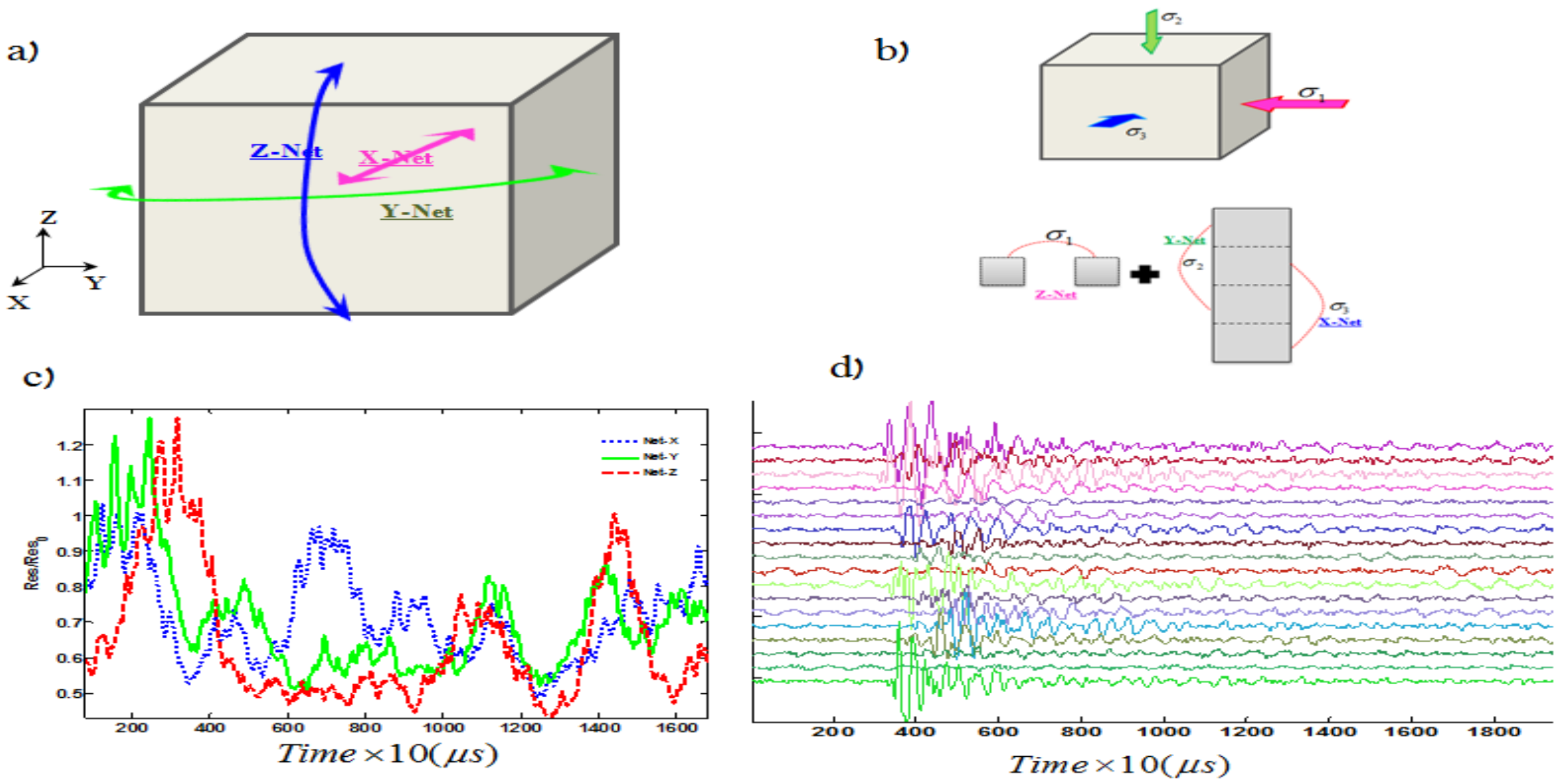

**Figure S.8 (a-b) An** illustration of sub-networks in X,Y and Z ; we use each layer of the network as an independent network. Parallel faces from a cubic sample form sub-networks of acoustic-sensors (or layer X,Y and Z) . (c) An example of R-profiles in each layer. (d) A typical recorded event through 18 transducers in ~180µs .

In Fig.S.7 –S.8, we have shown the collapsing of the best-strong events (i.e., $\{R_i^I\}_{i=1,2,3} > 1$) in local network parameter spaces : $\frac{1}{\log < B.C >_{x,y,z}}$ - $\bar{\lambda}_{\max_{x,y,z}}$. While the events are the identical events, Z-networks ($\sigma_1$ sensors) exhibits –on average- a long tail toward high value of $\frac{1}{\log < B.C >}$. Hence, the X layer ($\sigma_3$ direction) slightly hold higher $\bar{\lambda}_{\max}$.

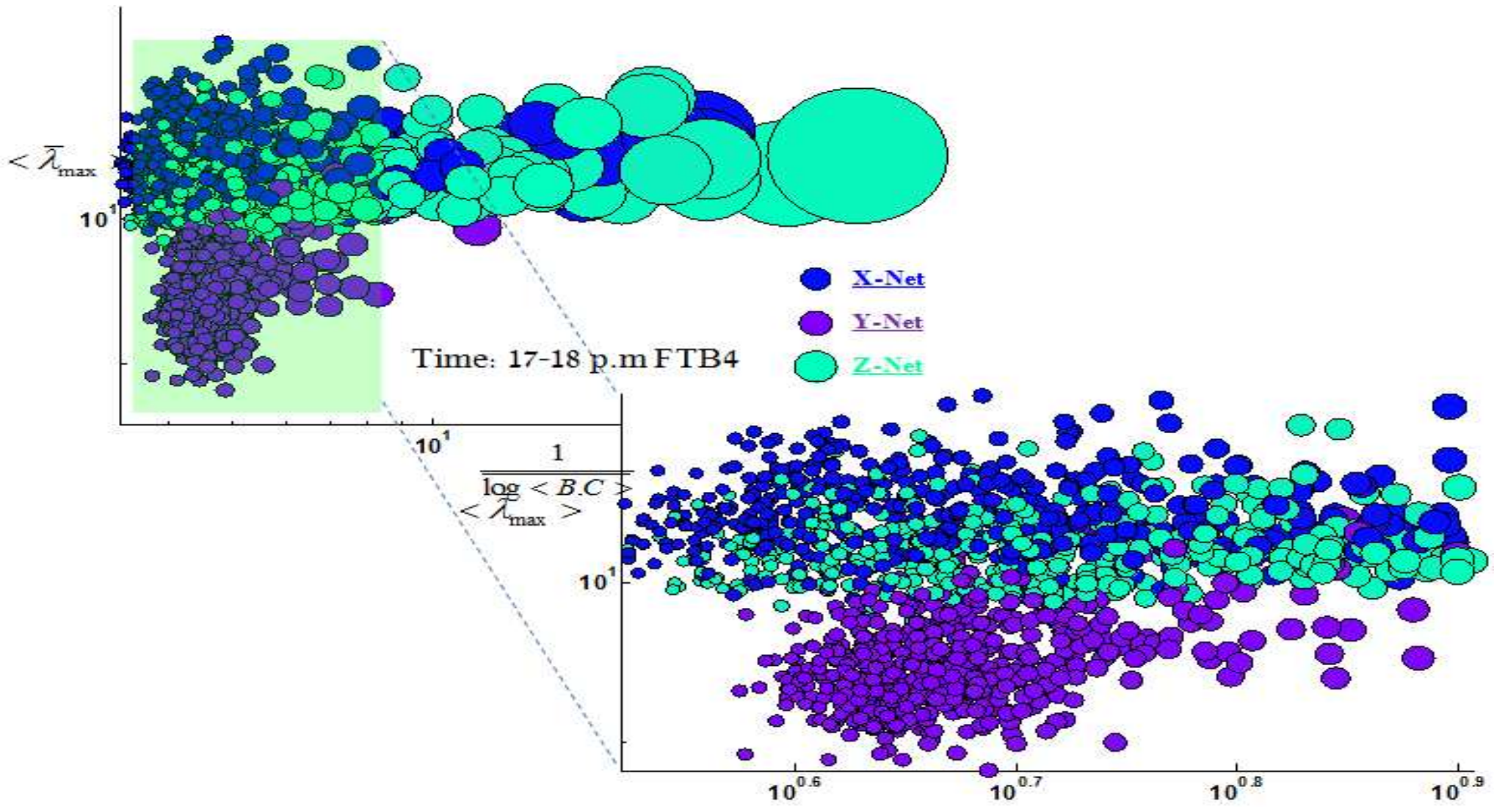

**Figure S.9** $\bar{\lambda}_{max.} - \overline{\log < B.C >}$ phase space for an hour of the FTB4 experiment. The size of the circles are proportional with the maximum of $\overline{\overline{\frac{1}{\log < B.C >}}}$. X and Z layers hold a high activity of the mapped events (corresponding to bigger $\overline{\overline{\frac{1}{\log < B.C >}}}$) with smoother trend in this phase diagram. Since events are not limited to a particular time or position, we interpret that the probability of growing micro-cracks in X-Z plane (minimum-max stress plane) is higher than other possibilities.

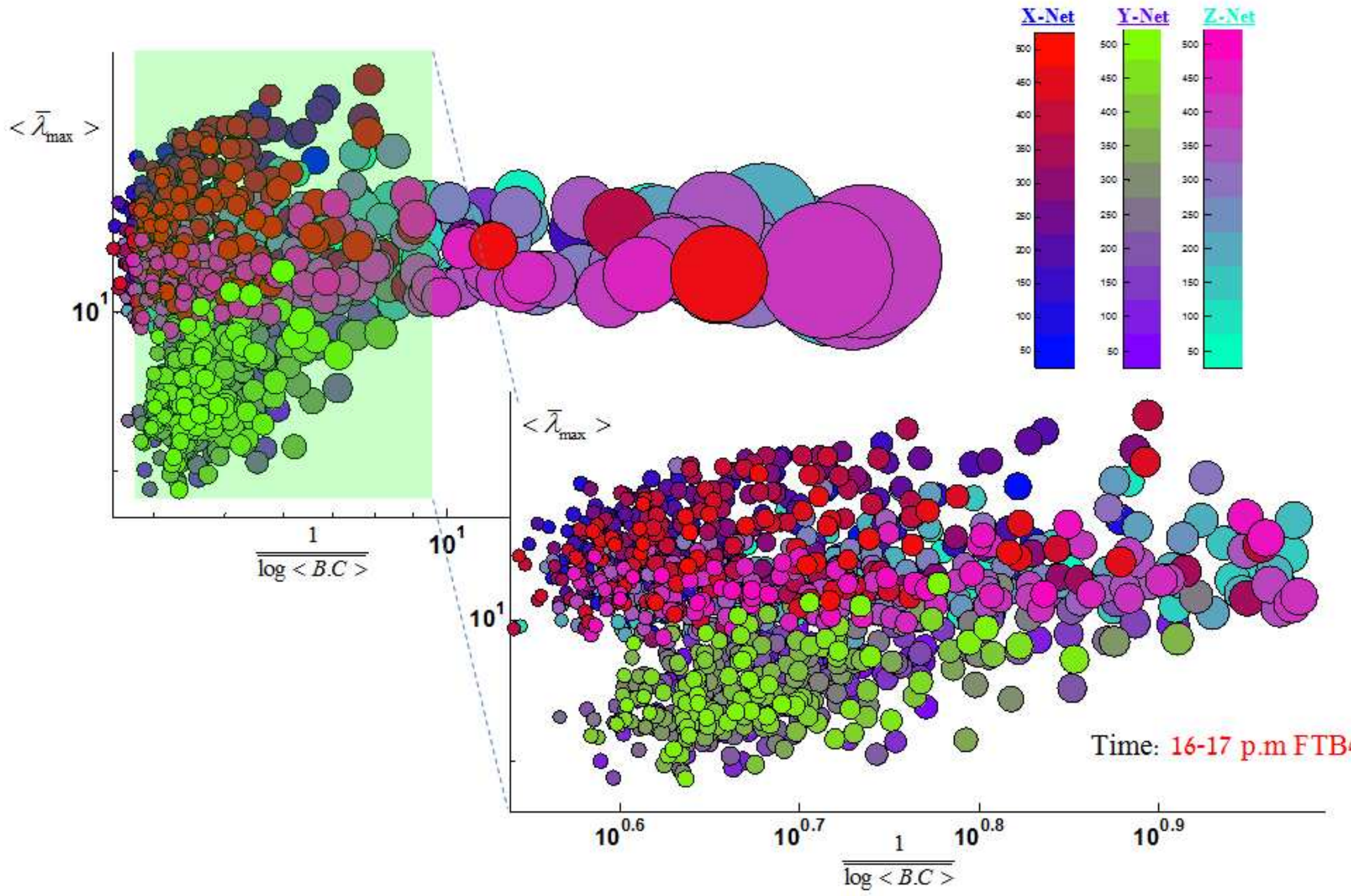

Figure S.10 $\bar{\lambda}_{\max.} - \overline{\log < B.C >}$ phase space for an hour of the FTB4 experiment. Considering a power law such as $\bar{\lambda}_{\max.} \sim \overline{\log < B.C >}^{-\chi_i}$ ,we have : $\chi_y > \chi_x > \chi_z$ . The size of the circles is proportional to the maximum of $\frac{1}{\overline{\log < B.C >}}$.

Moreover, events in the $(R_1^I, R_2^I, R_3^I)$ parameter space do not show a clear relation (scaling) while the totally diverse distribution in the aforementioned phase spaces is collapsed in a universal power law when we use $J_1 = R_1^I + R_2^I + R_3^I$ and $J_2 = R_1^I R_2^I + R_2^I R_3^I + R_1^I R_3^I$ (Fig.S9-S.10-S11). As we pointed out ,the maximum value of R-profile (at the onset of the fast-slip regime) is interpreted as an index to the maximum dynamic strength in sub-micron rupture scale. Then, maximum R values in each layer are assumed as a principle component of a deformation's attributes. With this interpretation, we used the first and second invariants ( $J_1 = R_1^I + R_2^I + R_3^I$ and $J_2 = R_1^I R_2^I + R_2^I R_3^I + R_1^I R_3^I$ ) to speculate about possible failure criterion. This procedure reveals a very close power law coefficient to previously reported similar criterion by Reches (3) on different rock-samples.

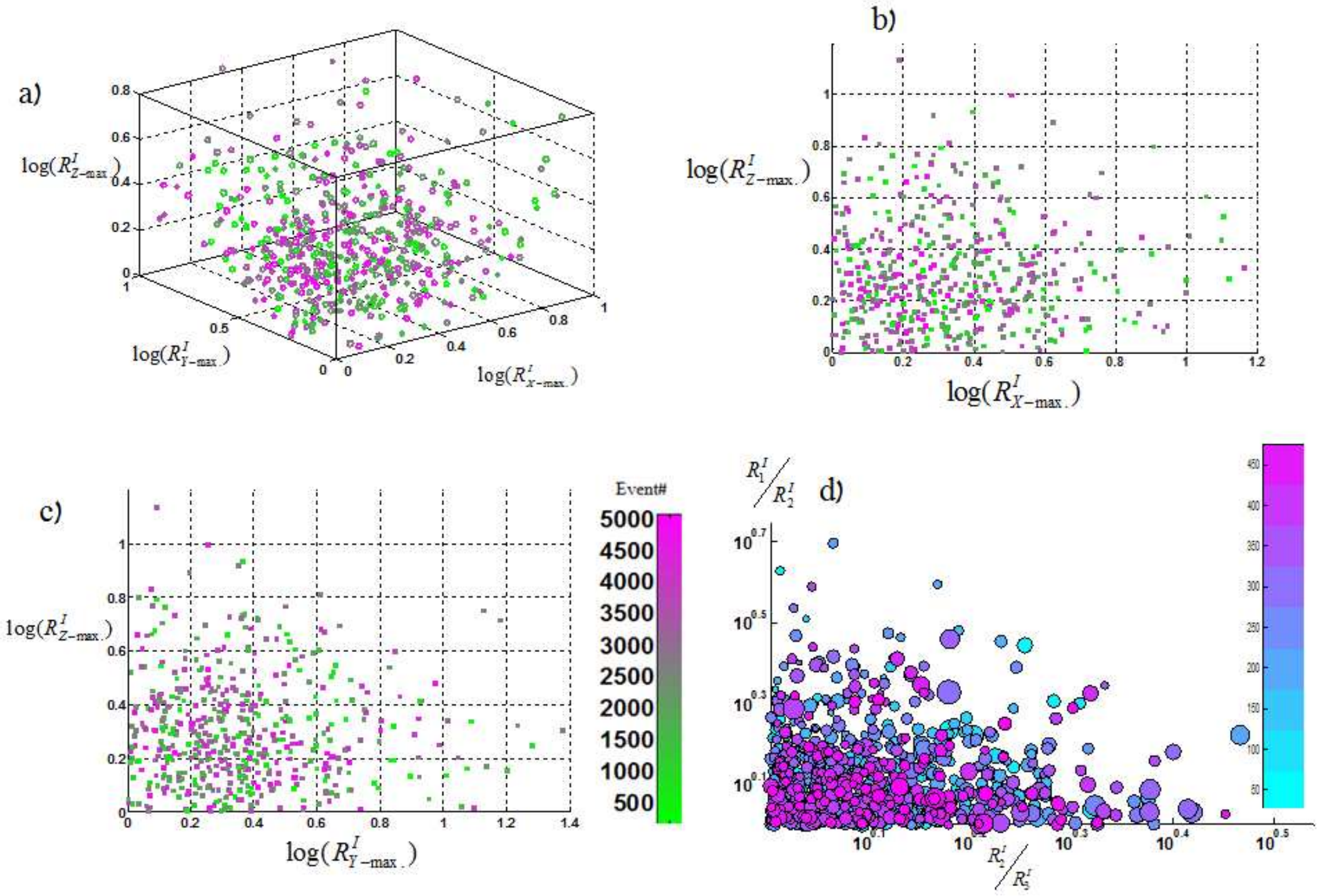

**Figure S.11** Precursor events in **the** $(R^I_{x-\max}, R^I_{y-\max}, R^I_{z-\max})$ parameter space do not show a clear relation between the maximum resistivity (or strength) of the events in the main directions. However, interestingly, the totally diverse distributions in the aforementioned phase spaces are collapsed in a universal power law while we use $J_1 = R^I_1 + R^I_2 + R^I_3$ and $J_2 = R^I_1 R^I_2 + R^I_2 R^I_3 + R^I_1 R^I_3$ .

Likewise, in Fig.S.10 (a and c ) and Fig.S.11a, with defining a similar parameter to octahedral shear stress $\Upsilon_R = \sqrt{(R^I_1 - R^I_2)^2 + (R^I_2 - R^I_3)^2 + (R^I_3 - R^I_1)^2}$ ,we could obtain the parameter spaces shown in Mogi (13) and Haimson (14). Now, with the interpretation of $\log \Upsilon_R \propto \beta \log R^I_1$ (Fig.S.10e , Fig.S.11a) as the local failure characteristics and considering $\log \Upsilon_R \equiv \tau, \log R^I_1 \equiv \sigma$ ,we reach the **similar -Coulomb law:** $\tau \propto \beta\sigma$ (also see last section for analysis of simpler tests 'events from saw-cut and rough fault of Westerly Granite rock samples).

Next, we find that the identical events are not classified in the similar way in the local and global sub-network/network spaces (Fig.S.11a-b), indicating the role of intra-links between layers in global networks. The (Eq.3-Methods part) of Q-profiles (of global network) is positive, showing a left-hand asymmetric shape of the pulses. Ruptures with higher energy (smaller $\overline{Q}$**)** hold a relatively more asymmetric shape (Fig.S.12). With approximation, an asymptotic curve may be fitted on global skewness ($\Sigma_g$)-$J_2$phase space, indicating that larger $J_2$ or $J_1$ leads to an average (or medium) skewness of Q-profiles (Fig.S.12d). Considering that larger $J_1$possibly results in longer waveforms and the relation among pulse duration and skewness in crackling noise systems (see Fig 28 in (15)), we conjecture that the results are similar to the aforesaid figure: longer waveforms approach smaller leftward asymmetry.

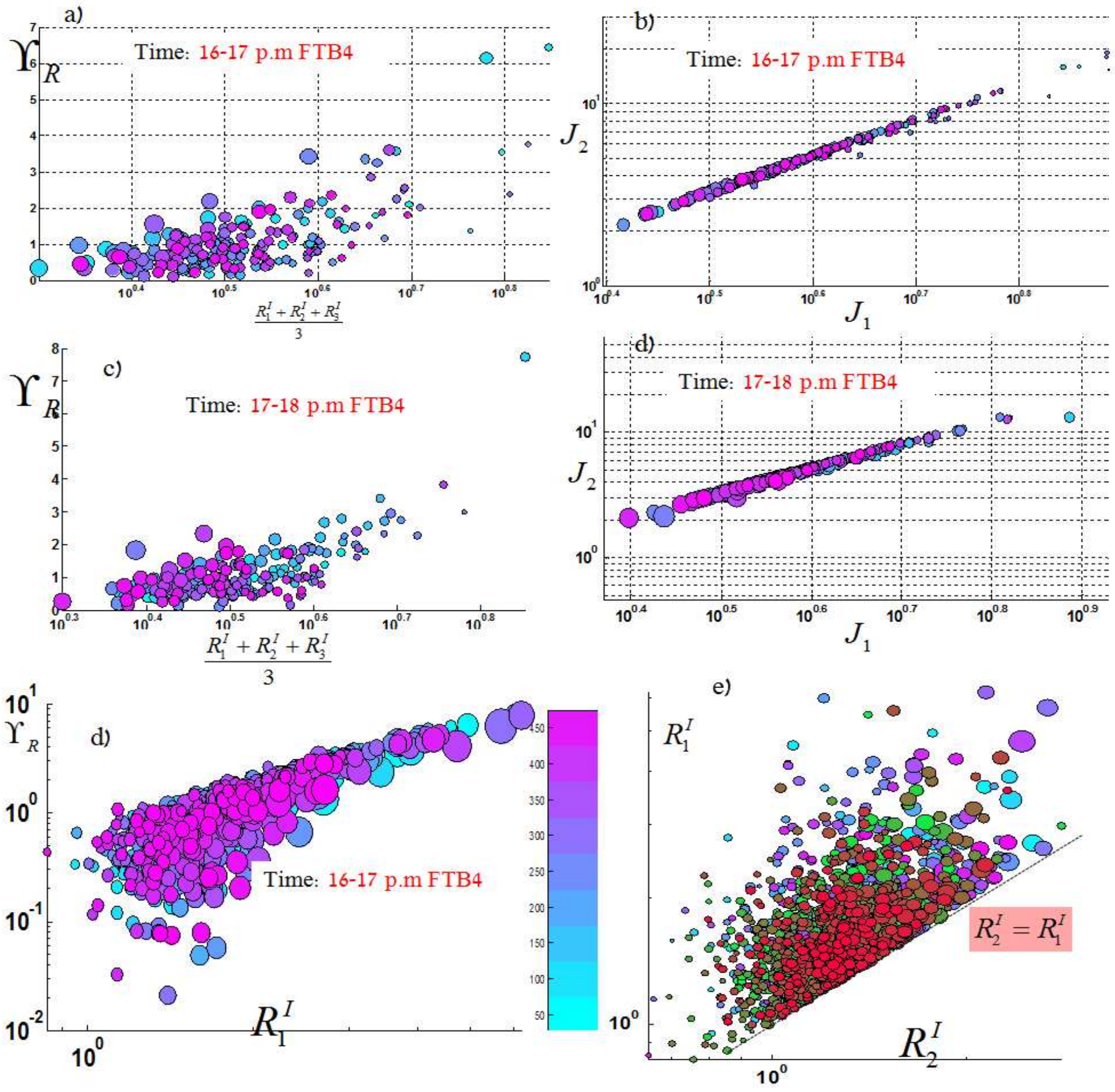

**Figure S.12 Failure criterion with minimum modularity of sub-networks .** (a,c) Collapsing ~520 best-strong events from the FTB4 tests in $\Upsilon_R$ - $R_1 + R_2 + R_3 / 3$ parameter space from two hours of the experiment, where $\Upsilon_R = \sqrt{(R_1^I - R_2^I)^2 + (R_2^I - R_3^I)^2 + (R_3^I - R_1^I)^2}$ (similar to octahedral shear stress-also see (3) and (13-14) for the similar scaling over final failures) . (b,d) Phase diagrams in $J_1 = R_1^I + R_2^I + R_3^I$ and $J_2 = R_1^I R_2^I + R_2^I R_3^I + R_1^I R_3^I$ . A power law function satisfies an excellent collapsing of events in the $J_1$-$J_2$ phase diagram as: $J_2 \propto J_1^b, b \approx 2.6$. (e) $\Upsilon_R$ - $R_1^I$ (with $\Upsilon_R \propto R_1^{\beta}$) shows that most of the events are collapsed in an interval with one order of magnitude difference on the $\Upsilon_R$ axis. This trend slightly changes after a threshold level of $R_1$ (on average, decreasing $\beta$) , comparable with panels a and c. The size of circles in a-e are proportional to $R_2^I$. (f) $R_2^I - R_1^I$ phase

diagram for two successive hours of the experiment. The size of circles is proportional to $R_3^I$. See also section 5 for more analysis of the parameter spaces in the simpler experiments.

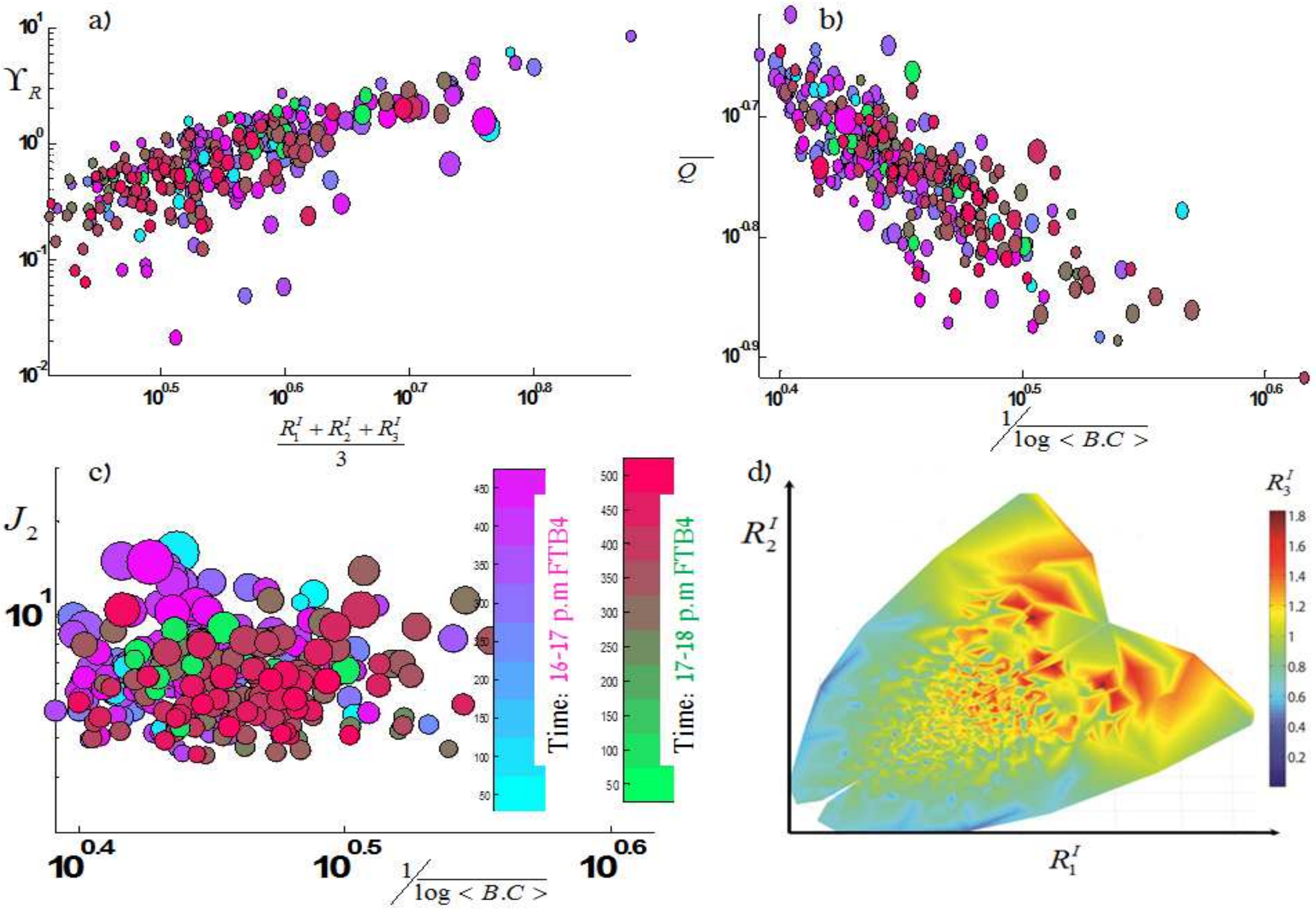

**Figure S.13.** Studies on network phase spaces. We have shown the events from two hours of the experiment (before main stress drop) in local (a) ,global(b) and global-local phase diagrams. The size of the circles are proportional with the $R_2^I$ in that occurred event. While (a) and (b) confirm the scaling of attributes,(c) does not show such a correlation. Also, note that the identical events are classified in different spots in (a) and (b), signifying different classification of rupture fronts in pure local and global sub-networks/networks phase spaces. The size of circles is proportional to to $R_2^I$. (d)We have shown a collection of planes on the triple points of max-R values for events occurred ~an hour before final failure (i.e., locus of local-yield surfaces).

**Relation between Global and Sub-Networks (independent layers)-** Next, using the simplified scaling relations, we infer the possible relation of fracture energy (inducing the duration of the second phase in Q-profiles) and local sub-networks (i.e., sub-micron/micron or precursor fronts' failure criterion). To proceed, the upper bound of the $J_2 - \Sigma_g$ phase diagram is approximated with $J_2 = \exp(-(\Sigma_g - \Sigma^c_g))$ and then replacing with $\Sigma_g \propto \overline{Q}^{-l}$, It leads to $J_2 \propto \exp(-\overline{Q}^{-l})$ for the upper bound (or approaching from the top). We further assume that the power-law coefficient $l$ is identical for both TTT and CTT. In addition, due to shorter generic phases of TTT (while we average over the same time interval), we expect $\overline{Q}_{CTT} > \overline{Q}_{TTT}$ which yields $J_2^{CTT} > J_2^{TTT}$ for $\Sigma_g > \Sigma^c_g$ (and vice versa). We conclude that higher than a threshold level, higher $J_1$ (or $J_2$) couples with a higher fracture energy then resulting in a slightly lower $\beta$ in $\Upsilon_R \propto R_1^{\beta}$. Then, we connected global networks to local sub-networks' metrics.

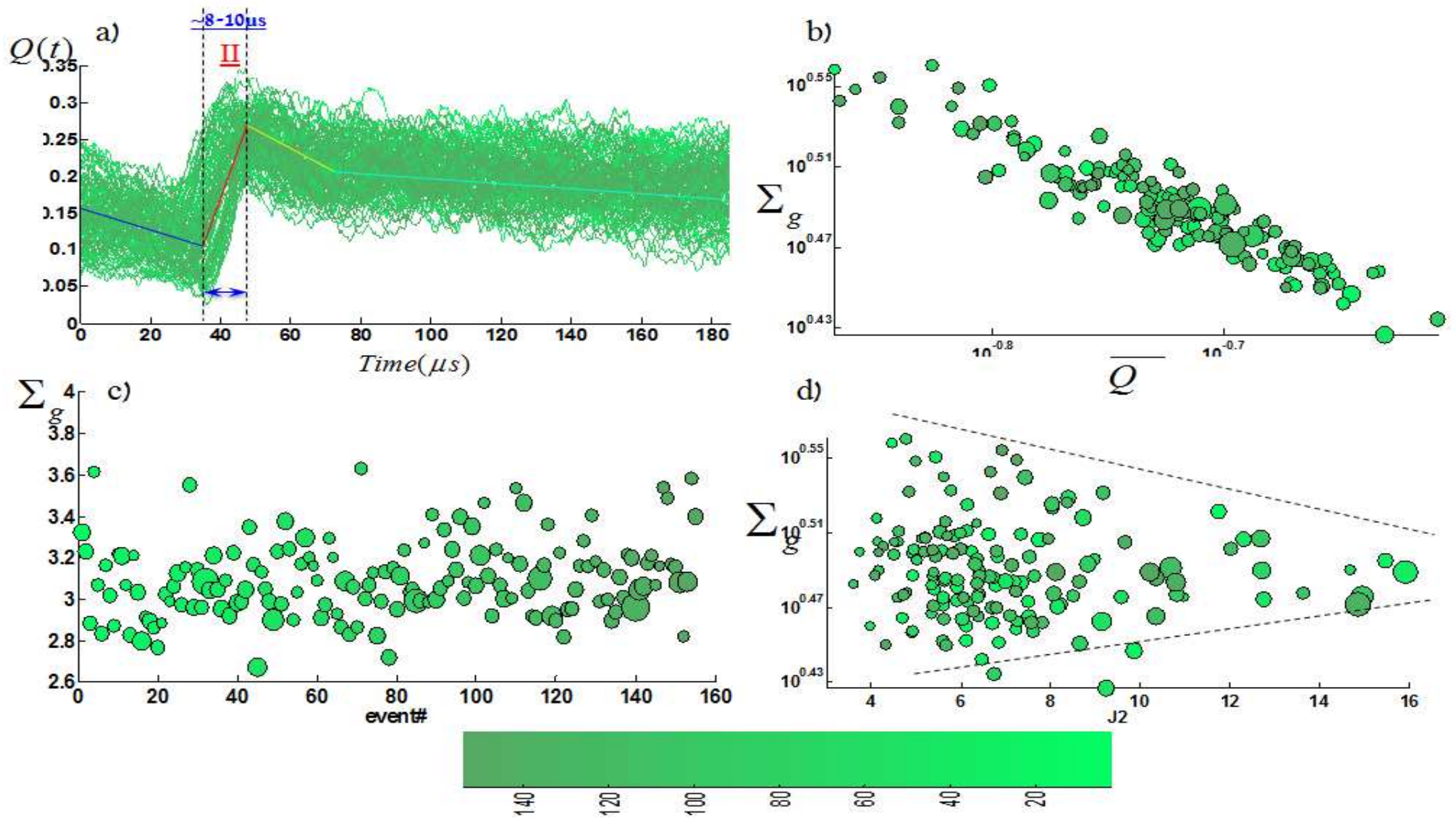

**Figure S.14.** The average skewness of Q-profiles from global networks. (a)A collection of ~160 events in TTT (FTB4) experiment , schematically representative of generic phases of Q-profiles. (b) $\Sigma_g \propto \overline{Q}^{-l}$ : skewness versus mean modularity .(c) The evolution of skewness for the sequence of events .(d) Global skewness versus local metric of maximum R values from sub-networks. The dashed lines show the upper and lower bounds/trends of events. The size of circles is proportional to to $R_2^l$.

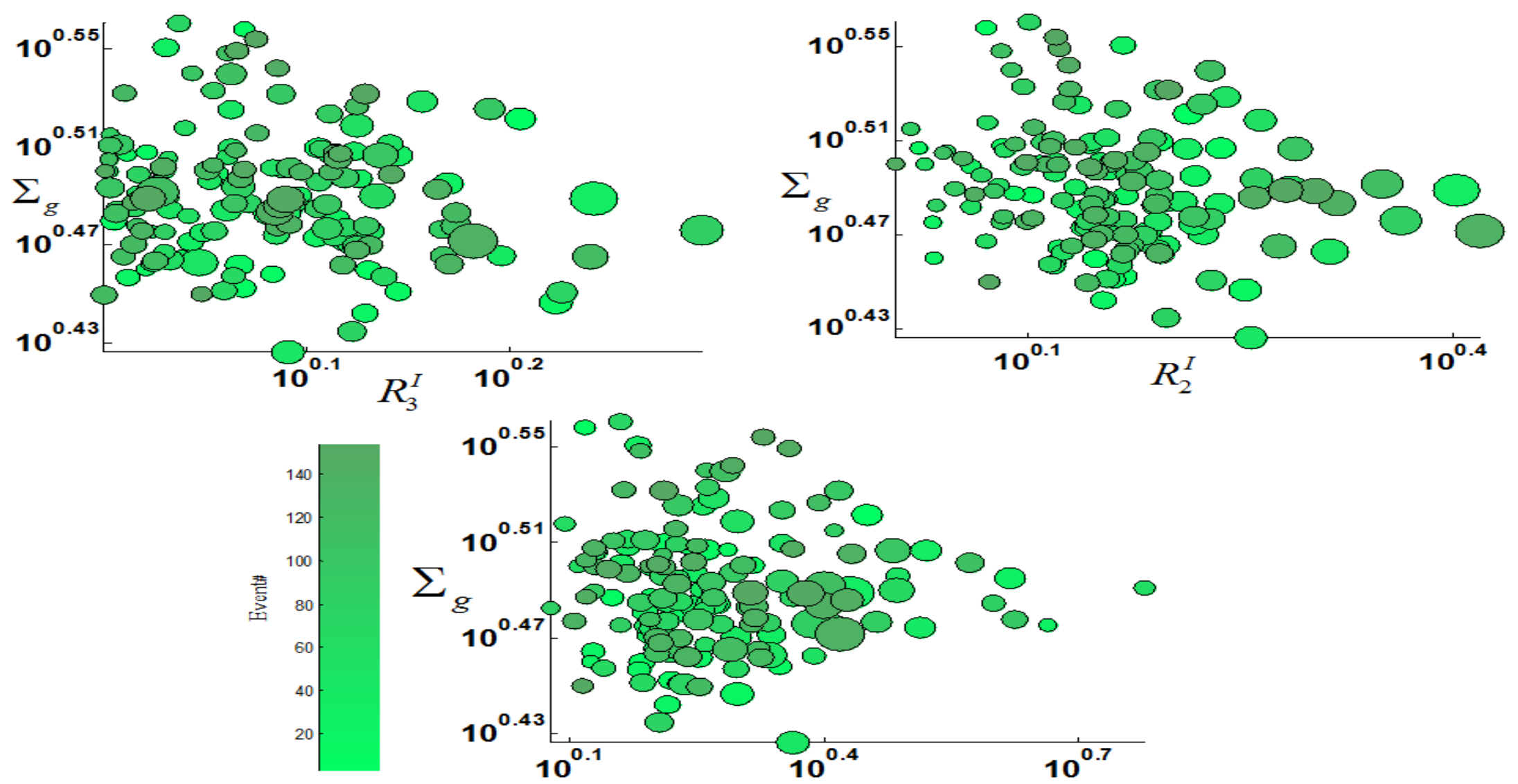

**Figure S.15** The average skewness of Q-profiles ($\Sigma_g$) versus max. R in each layer.

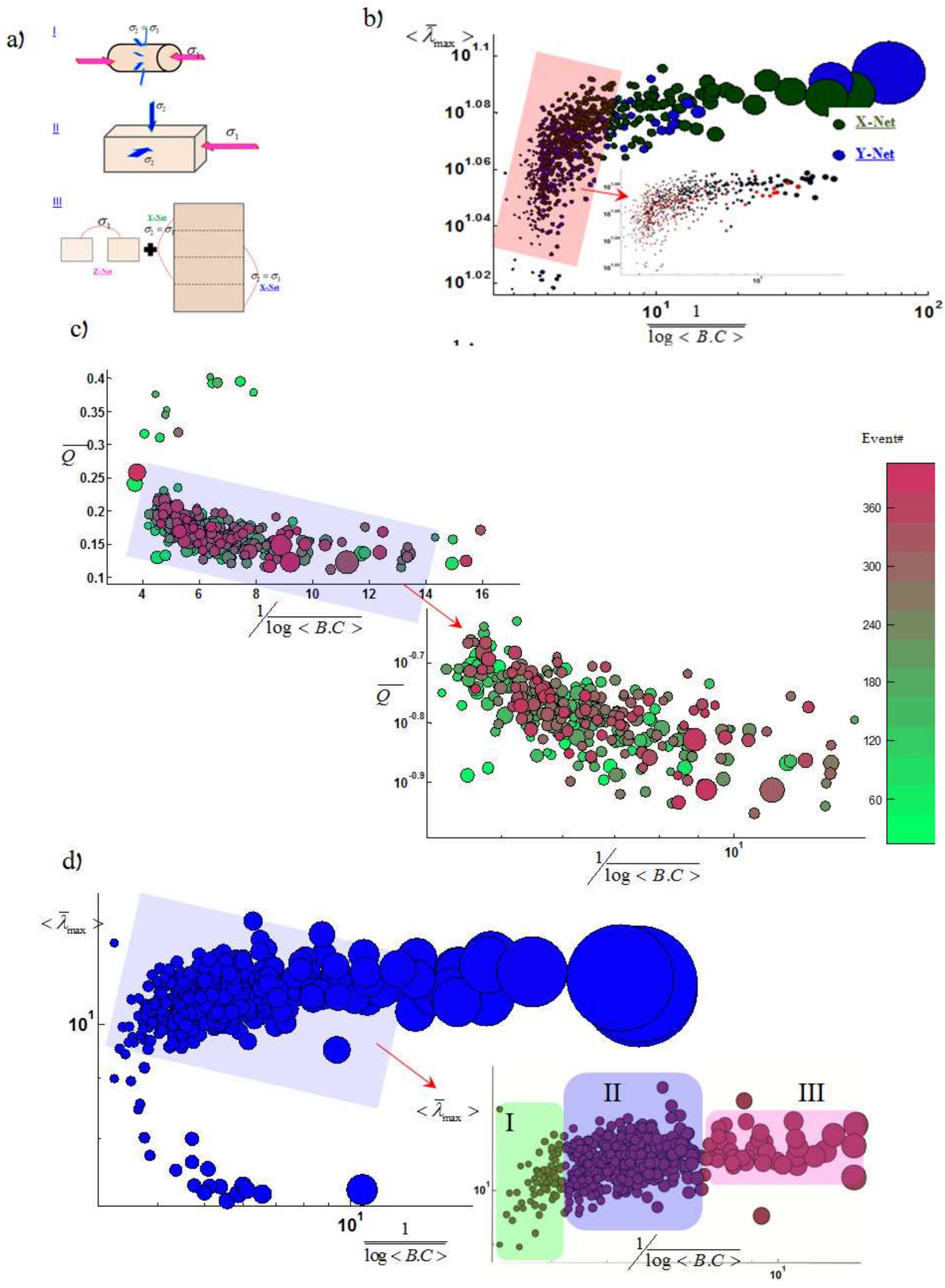

**Figure S.16.** Global and local Networks on CTT. (a) mapping a cylindrical sample to the flatten surfaces . With this technique, we analysis X and Y sub-networks while Z-sub networks are ignored (due to the lack of sensors in corresponding planes) .(b) X and Y layers' phase space shows a much more uniform distribution of events rather than distinguished cluster in TTT's sub-network. (c) The same events from panel (b) in the global networks' metric (all events satisfy $\{R_i^I\}_{i=1,2} > 1$) .(d) ~600

events <u>without considering</u> $\{R_i^I\}_{i=1,2} > 1$ from the CTT test unravel a universal picture of the recorded acoustic events with three main trends. The classified events in the inset of the panel (d) indicates the three main trends of $\bar{\lambda}_{\max.} \sim \overline{\log < B.C >}^{-\chi}$ based on the best fitted power law per each cluster. Most of events allocate the second cluster .For FTB4 ,the trend of this phase plane is different in a way that the first trend is nearly missing and the slop of clusters are steeper (see Fig.S.5b).

## 4. Further Experimental Results on TTTs

In this section, we present two more experimental results regarding true triaxial experiments to support our results on FTB4 and the proposed approaches. Repeating the same experimental procedure on the same rock material, while changing the intermediate and the minimum driving stress field. For FTB3 and FTB2, the configurations of the driving stress fields are as follows: **FTB3** $\sigma_3 = 10, \sigma_2 = 50$ and **FTB2:** $\sigma_3 = 10, \sigma_2 = 20$. In Fig.s.15, we have compared the duration of the fast slip phase for three TTTs. The duration of the second evolutionary phase with considering identical number of nodes for all cases for FTB3, FTB4 and FTB2 are ~20 µs ,~10 µs and ~10 µs, respectively. To explain the longer duration of FTB3 (which approaches to the CTT because is not real triaxial test due to imbalance of forces in this experiment) ,we compare the $\frac{1}{\overline{\log < B.C >_{x,y,z}}}$ - $\bar{\lambda}_{\max_{x,y,z}}$. over **FTB4** , **FTB3** and **FTB2** (Fig.S.16). With regard to approximated power law coefficient in the phase diagram $\bar{\lambda}_{\max.} \sim \overline{\log < B.C >}^{-\chi_i}$ , we have for FTB3: $\chi_x \gg \chi_y \geq \chi_z$ (on average), while for other cases $\chi_y \gg \chi_x > \chi_z$ . We logically infer that the duration of the second phase is related to the trends of layers in the phase diagram. Thus, micro-cracks with preferentiality to grow in the xz plane imprint a shorter duration and then reach fast "death". The main common feature of three TTTs are the distinguished cluster of Y sub-networks, which are always below X and Z layers; this could be assumed as a crucial role of the intermediate stress. Indeed, with approaching to peak stress (in z-direction) , the power law coefficient of Y-layer slightly increases.

Outstandingly, for the CTT cases the X and Y layers are nearly indistinguishable (Fig.S.14),stating a nearly *uniform* propagation of fracture ; In FTB3 case, higher power law coefficient of X-sub networks and smaller coefficient of Y-sub networks induced regular crack growth (thus *probability of propagation of micro-planes in Y-direction is higher:* <u>regular cracks</u>). We conclude that the probability of finding short-events and somehow finding anti-cracks (in terms of their generic phases) occurring in XZ plane is higher than YZ plane; then:

$$\begin{cases} \tau_{II}^{CTT} \approx cte & \sigma_2 = \sigma_3, \\ \tau_{II}^{TTT} \leq \tau_{II}^{CTT} & \sigma_2 \neq \sigma_3 \end{cases}.$$

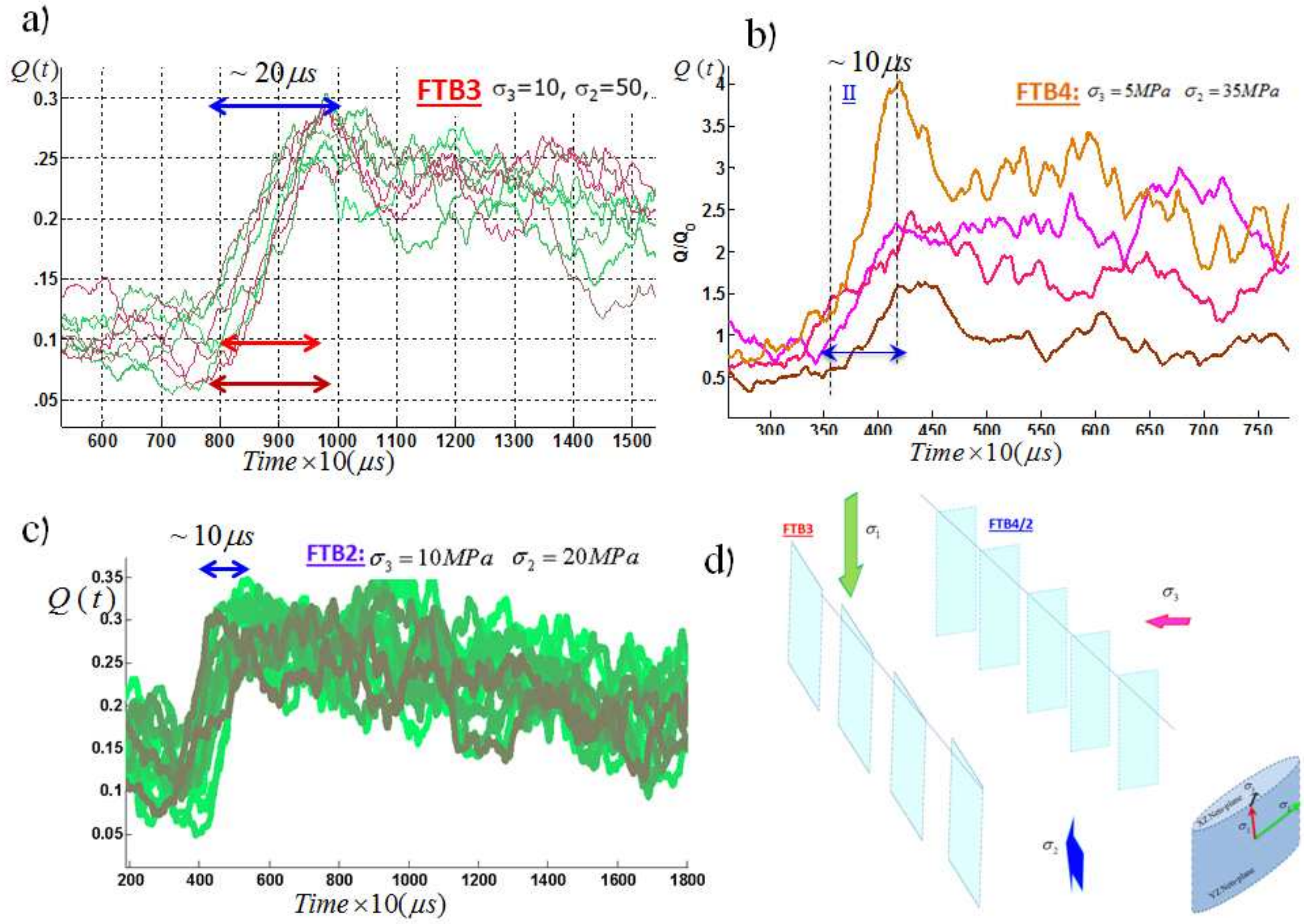

**Figure S.17. Time characteristic of fast slip phases for three different true triaxial tests.** (a) FTB3 $\sigma_3 = 10, \sigma_2 = 50$ which is 3D stress field but not true triaxial.' (b)FTB4: $\sigma_3 = 5, \sigma_2 = 35$ and (c) FTB2: $\sigma_3 = 10, \sigma_2 = 20$. (d) Based on the network phase spaces, a schematic presentation of fractures embedded in two planes has been shown. We assume that Z-net is driving the stress field (sigma 1 is the driving force). Events in XZ planes (quasi-anti cracks) hold a shorter fast-slip (annealing) phase while rupture fronts in YZ plane allocate longer duration. Generally, fast-ruptures with high energy grow in YZ plane .For conventional cylindrical testes, XZ plane is approximately identical with YZ plane.

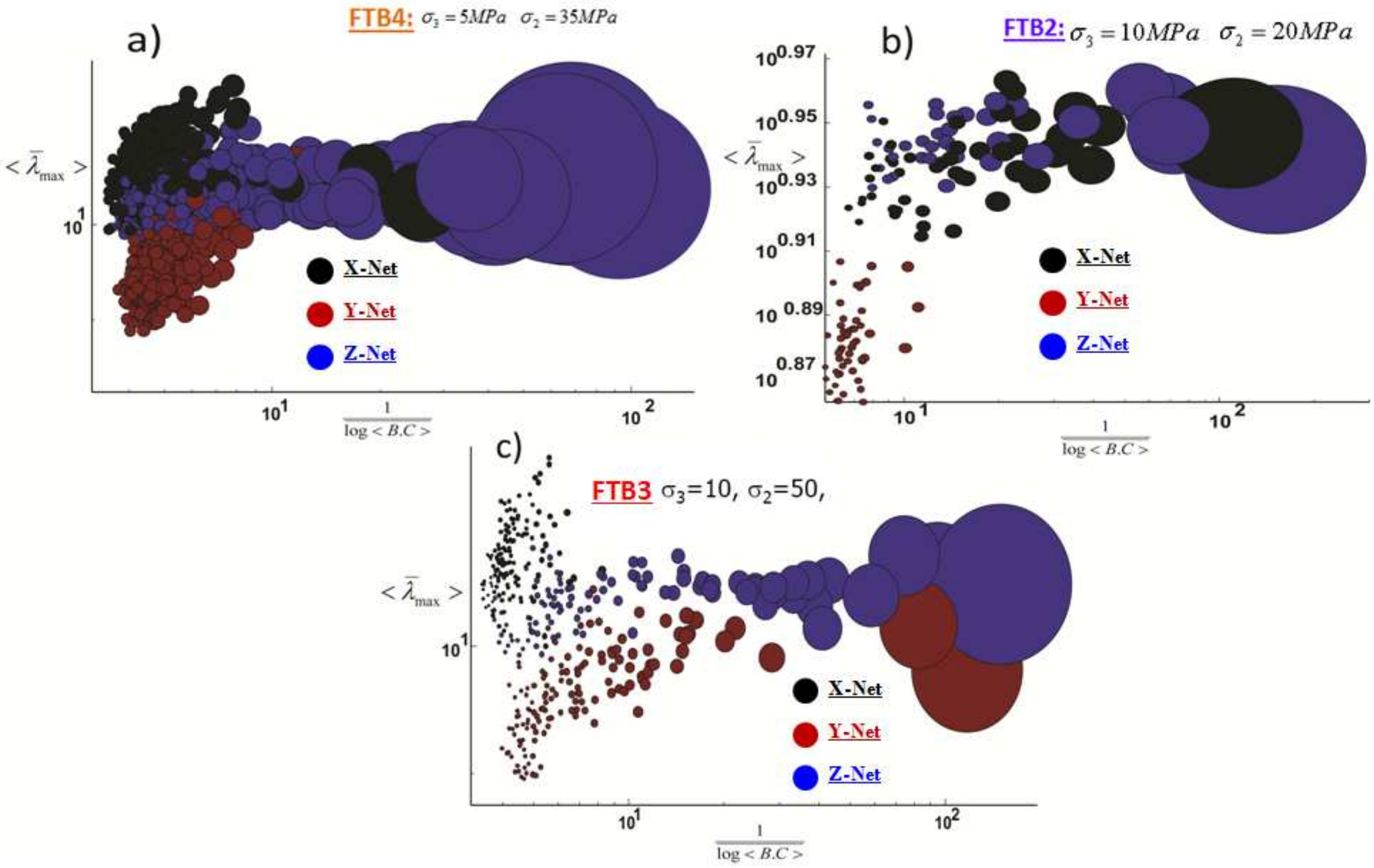

Figure S.18. | **Study of sub-networks** . Phase diagram of $\bar{\lambda}_{\max}^{x,y,z} - \overline{\log < B.C >_{x,y,z}}^{-1}$ for **(a) FTB4:** $\sigma_3 = 5, \sigma_2 = 35MPa$ **(b) FTB2** $\sigma_3 = 10, \sigma_2 = 20MPa$ **;.**The trends of FTB2 and FTB4' phase spaces are similar ,which is compatible with the similarity of the duration of the fast-slip phase as has been shown in Fig.2a and inset . **(c) FTB3** $\sigma_3 = 10, \sigma_2 = 50MPa$ which is a non-true traixial test**.** Considering a power law such as $\bar{\lambda}_{\max.} \sim \overline{\log < B.C >}^{-\chi_i}$ ,we have-on average- : $\chi_x > \chi_y > \chi_z$, representing a dominant propagation of rupture fronts in *yz* plane. The size of circles is proportional with the magnitude of $\overline{\log < B.C >_{x,y,z}}^{-1}$.

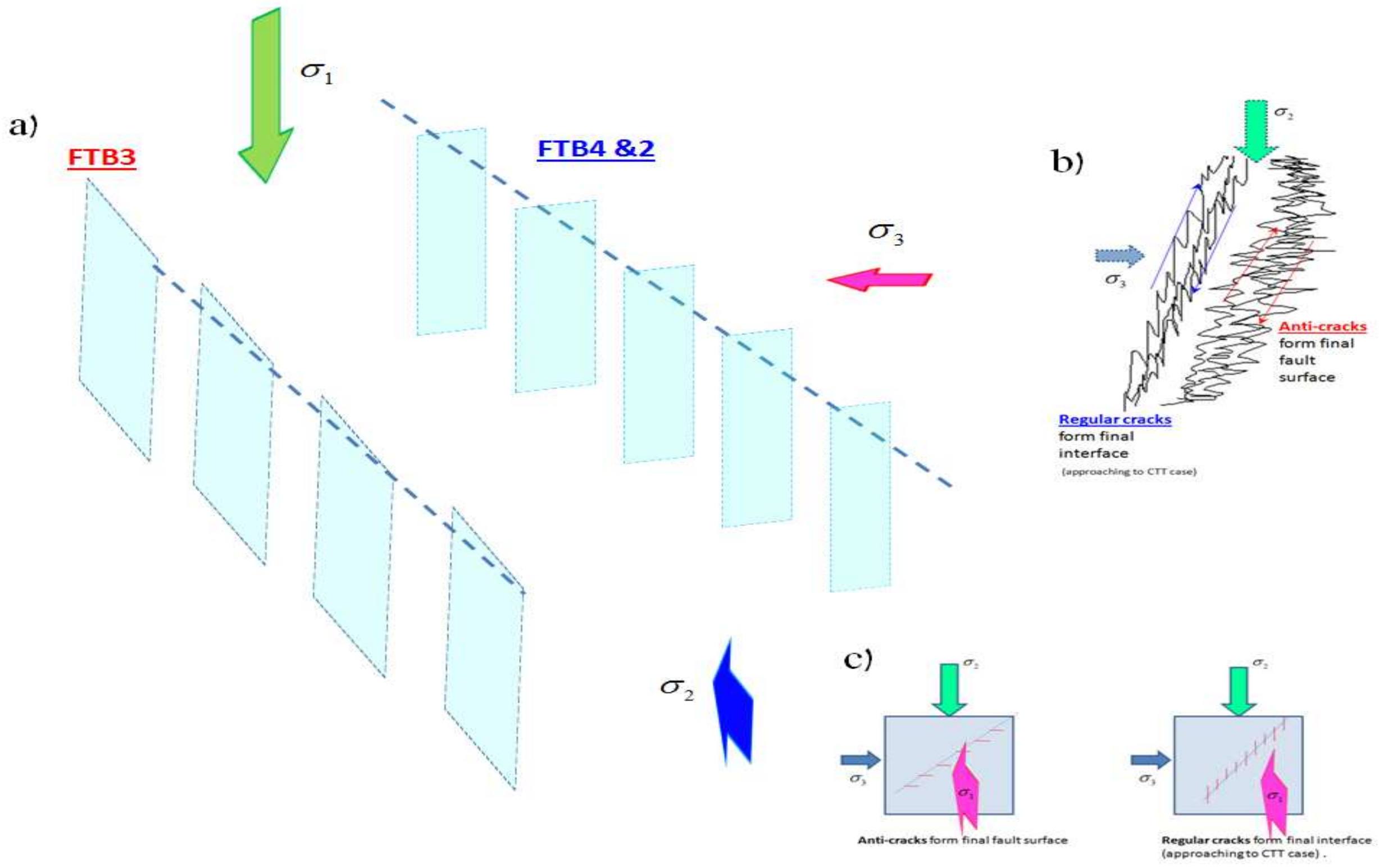

**Figure S.19**| schematic representations of two model-based regular and irregular cracks (anti-cracks) interactions to develop a self-organized nucleation of f a fault plane. The regular cracks and their interactions have been established through early CTTs and the theory of M-curve proposed in (16-7).

## 5. Discussion

To compare the TTT's sub-networks and global networks with CTT 's functional-networks' features , we use the following technique (considering that a cylinder can be seen an *n-gonal prism* where *n* approaches infinity). We flatten the cylindrical samples and,ignoring the curvature of surfaces, *n* planes plus two circles are produced. Since, in our acoustics set-up, there is no acoustic transducer in the top and bottom of the cylindrical samples, we ignore those two additional top and bottom circular planes (Fig.S.14a). With this mapping, assuming *n=4*, we reach an equivalent scenario to the flatten cubic (Fig.S.6b). Repeating the same procedure and with the same number of nodes in each layer (here two layers: X and Y), we find a different trend of X-Y sub-networks' phase diagrams (Fig.S.14b). Here ,events occupy homogeneously the phase space of $\overline{\lambda}_{\max}$ - $\overline{\log < B.C >}^{-1}$ ,while in the TTT results Y-networks (pre)dominantly reveal a separated cluster with more sensitivity to the control variable ($\overline{\log < B.C >}^{-1}$). In the scale of the global networks, the range of the networks 'parameters are broader than TTT's diagnostics which in turn induces more broad range of the shape of waveforms and then micron/sub-micron rupture regimes (Fig.S.14c &. Fig.S.11b). As a summary to the comparison, we showed that ruptures in the conventional cylindrical experiments may represent more abnormality which can be interpreted in terms of the quality of energy flow in to crack tip.

Next, to analyzes patterns of Multiplex networks (in this case independent sub-networks) ,we use the results of two-well studied experiments on smooth (saw-cut) and rough (natural fault) faults – embedded in the Westerly granite samples (7,18). Similar to the CTT-Sandstone test, we find that X-Y sub networks do not separate in different clusters and nearly all the patterns are uniform in our "standard" network phase space (Fig.S.18).

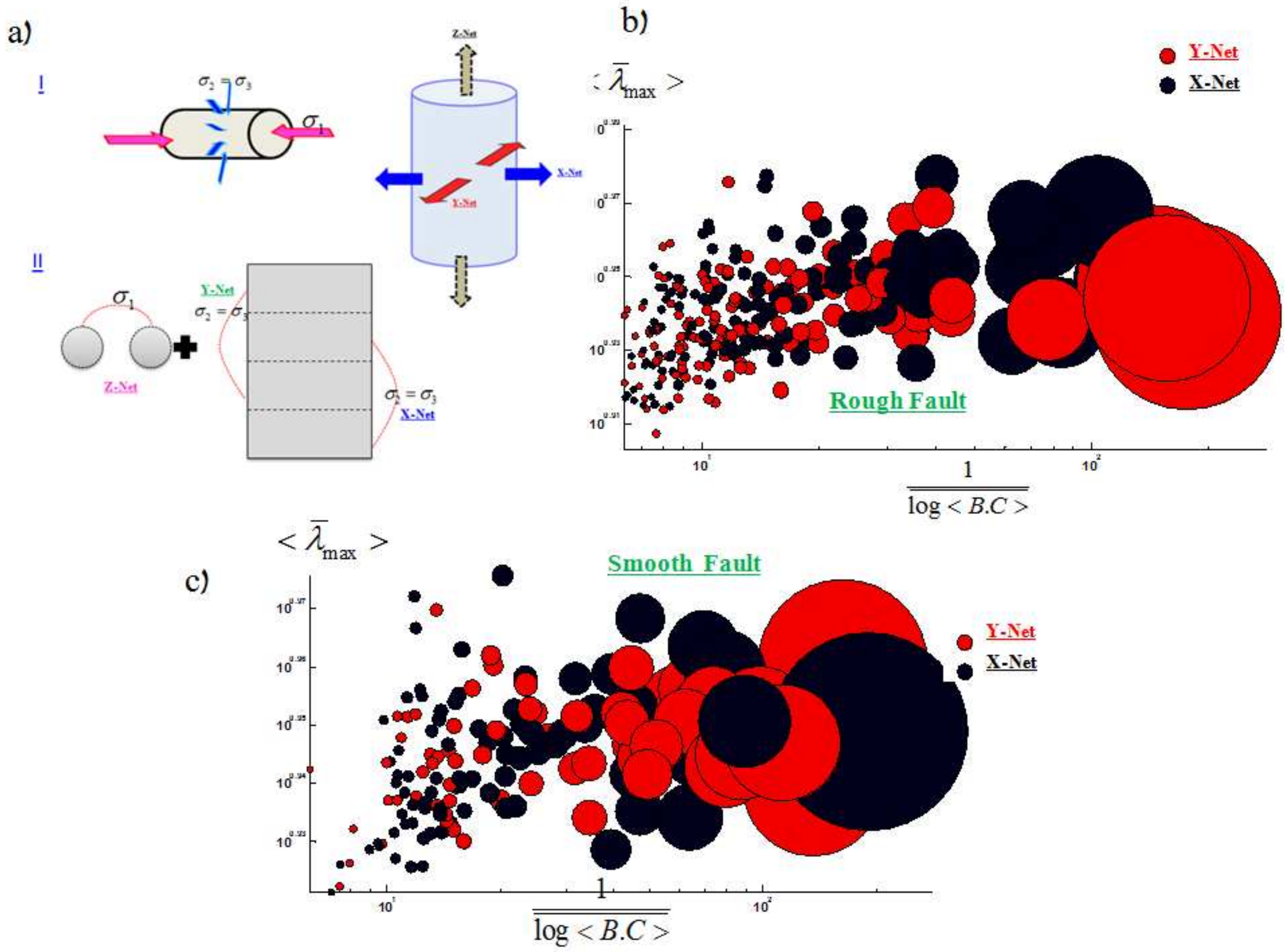

**Figure S.20**|(b) X-Y sub networks' phase space on events from the naturally made-rough fault (18) (c) Multiplex Networks and X-Y sub networks' phase space on events from the embedded saw-cut fault in~60 in Westerly Granite (18) . No significant separation or clusters can be recognized in the $\frac{1}{\overline{\log < B.C >_{x,y}}} - \overline{\lambda}_{\max_{x,y}}$ plane.

Using events from the saw-cut fault experiment (~100 events in three cycles of stick-slip cycles) and mapping the maximum and minimum R values from each event (from the first phase) ,we find a failure-criterion phase space ,similar to TTT (FTB4) . The results show three distinctive trends in $\Upsilon_R - R_{\max}$ (Figs.19a) : 1) $R_{\max} \to R_{\min}$ which results in an insensitive section of the phase space .2) $\Upsilon_R \sim R^{\beta}{}_{\max}, \beta > 1$ and (3) a few events occupy: $\Upsilon_R \sim R^{\beta}{}_{\max}, \beta \to 1$.

Interestingly, the trend obtained is similar to the recent results of precursor events' evolution in PMMA-frictional interfaces as has been reported in Fig2. of (19). Increasing the length of the crack in one direction is obviously related to the magnitude of R values in X or Y networks. Then (precursory) cracks with smaller length allocate the smaller $R_{\max}$ while increasing "shear stress" will not imprint a dramatic change in the crack length. However, approaching final crack length or main (big) stress drop (stick-slip) will happen in the third section of the phase space. With having this interpretation and the results of (19)-supporting our approach- we can interpret the results of Fig.S10d which share two main trends of Saw-cut fault experiment (i.e., the critical trend or the stage 3 is missing in TTT).

To prove that the observed short-phase Q-profiles are real anti-cracks, we compare typical anti-cracks' waveforms with TTT's Q-profiles. we set-up a global network on a multi-anvil test (accomplished at University College London) on Ge-Olivine samples. The procedure of the test was similar to (20) with confining pressure of 2 to 5 GPa and temperature between 800-1500K. In Fig.S.12, we have shown two typical waveforms and their corresponding Q-profiles (from the deformation phase at ~2GPa), demonstrating an encoded short duration of the second phase <10μs. Since one of the main mechanisms to faulting at high pressure tests is associated with anticrack development (21-22) ,we claim that the short duration of the second generic phase is due to abnormal cracking (the second phase of Q-profiles from samples under regular CTT lasts ~20-25 μs) .

Our observation of shorter rupture time and shorter fast-slip phase in multi-anvil test does coincide with Heidi Houston and Quentin Williams 's conclusions (23-24) that many deep events (deep-focus earthquakes) start up significantly faster than do intermediate or shallow events. In addition, our results from multi-anvil test confirmed the results of (24) that the entire rupture time of deep-earthquakes (with possible anti-cracks' origin) is about half of shallower events.

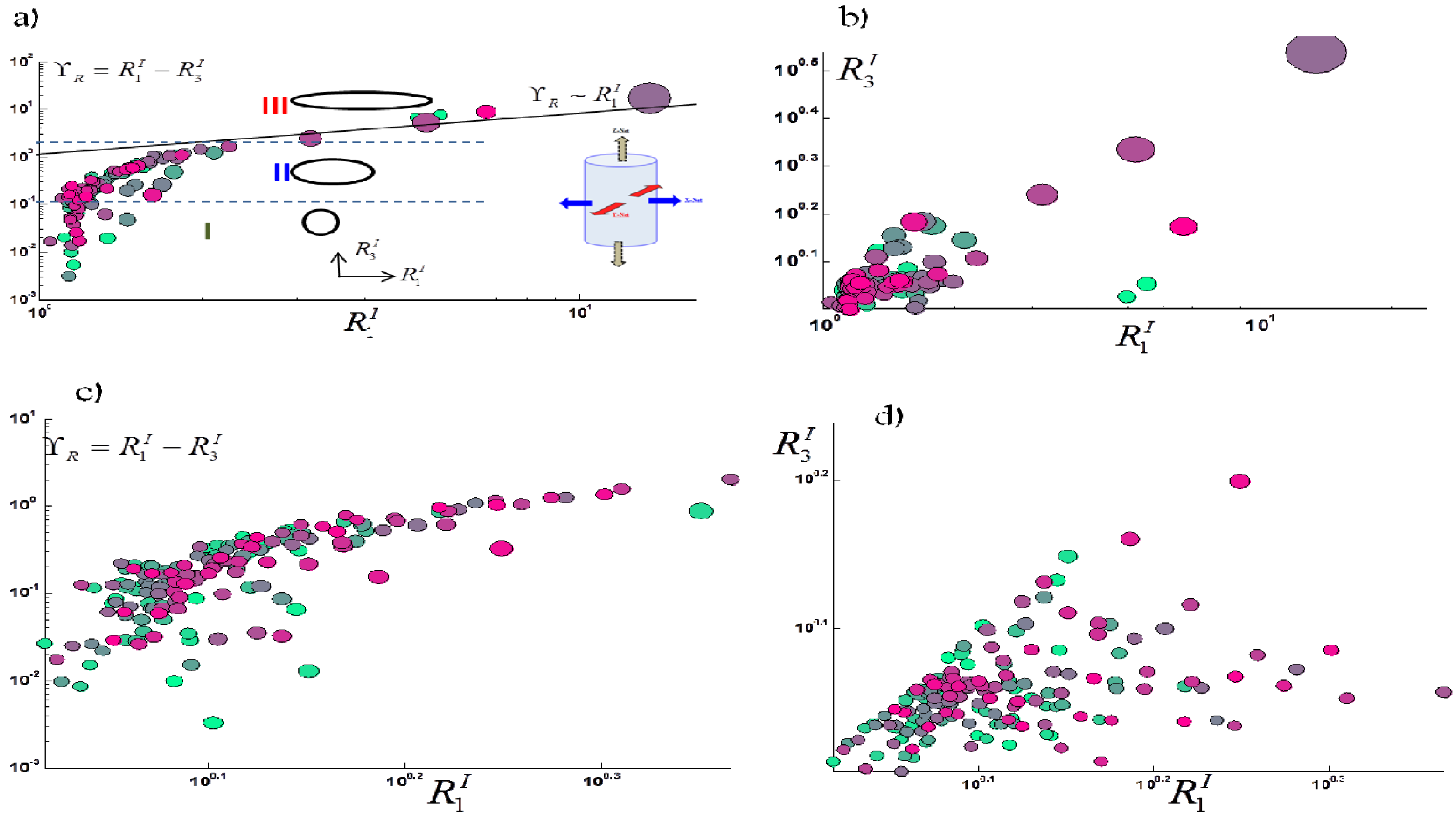

**Figure S.21| (a-b)** Maximum R values from X and Y sub-networks from a Smooth-fault and **(c-d)a** Rough fault**.** The color shows the events’ sequence and the size of circles is proportional to the minimum R value.

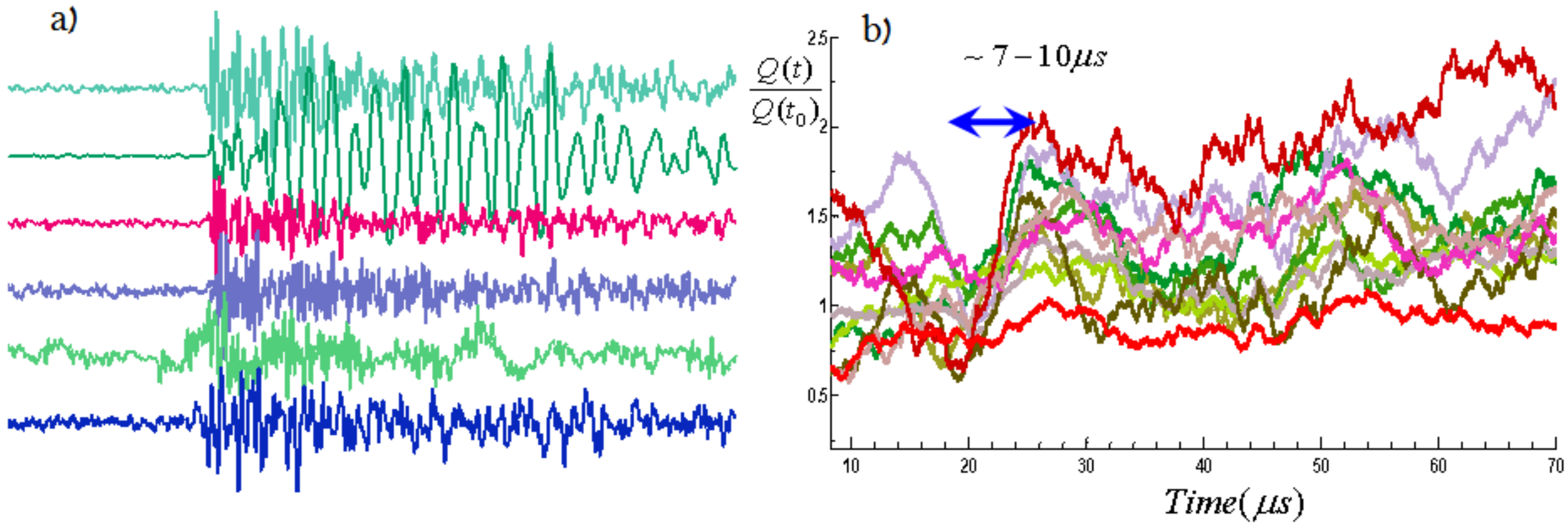

**Figure S.22| Typical waveforms proposed to have the origin of anti-cracks in a Multi-Anvil test;** We have shown typical waveforms from an event (a) from anti-crack faulting in Olivine samples and a collection of 10 events and their corresponding Q-profiles. The duration of the second phase is ~<10 µs. The regular duration of this phase for regular micro cracking is ~20-25 µs. This is a typical range for most of rocks with frequent $SiO_2$ (such as Granite, Basalt, and some Sandstone).

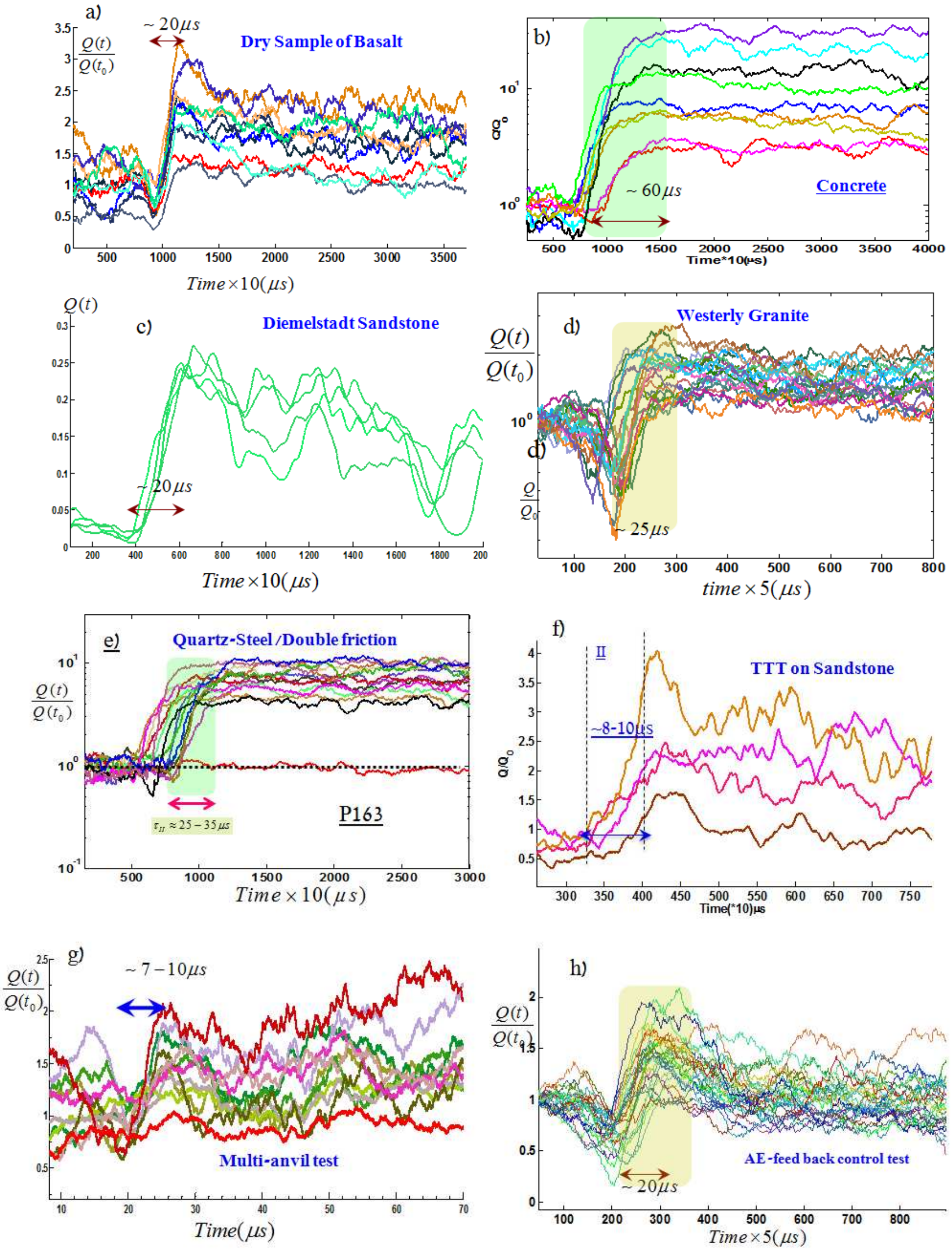

**Figure S.23| Summery of studied waveforms from 9-set of experiments on different brittle materials.** Our test included simple friction-test, conventional cylindrical triaxial tests, polyaxial tests, High pressure –High temperature, tests on frictional-gouge interfaces . We used Granite, Sandstone, Basalt, Olivine and Concrete samples.